\documentclass[reprint,amsmath,amssymb,prd, nofootinbib, floatfix, preprintnumbers, superscriptaddress]{revtex4-2}

\usepackage{graphicx}
\usepackage{dcolumn}
\usepackage{xcolor}
\usepackage{bm}
\usepackage{array}
\usepackage{hyperref}
\usepackage{lipsum}
\hypersetup{
     colorlinks   = true,
     citecolor    = blue
}
\usepackage{xurl}
\usepackage{lineno}

\newcommand{\IFCA}{\affiliation{Instituto de F\'isica de Cantabria (IFCA, UC-CSIC), Avenida de Los Castros s/n, 39005 Santander, Spain}}
\newcommand{\UVA}{\affiliation{University of Virginia, Department of Astronomy, Charlottesville, VA, 22904, USA}}
\newcommand{\KCL}{\affiliation{Department of Physics, King's College London, UK}}

\newcommand{\TDLI}{\affiliation{Tsung-Dao Lee Institute (TDLI), No.\ 1 Lisuo Road, 201210 Shanghai, China}}
\newcommand{\SJTU}{\affiliation{School of Physics and Astronomy, Shanghai Jiao Tong University, 800 Dongchuan Road, 200240 Shanghai, China}}

\begin{document}



\title{Axions in Andromeda: Searching for Minicluster - Neutron Star Encounters \\ with the Green Bank Telescope}

\author{Liam Walters}
\UVA

\author{Jordan E. Shroyer}
\UVA

\author{Madeleine Edenton}
\UVA

\author{Prakamya Agrawal}
\UVA


\author{Bradley R.\ Johnson}
\UVA

\author{Bradley J.\ Kavanagh}
\IFCA

\author{David J.\ E.\ Marsh}
\KCL


\author{Luca Visinelli}
\TDLI \SJTU

 


\preprint{APS/123-QED}
\date{\today}

\begin{abstract}
The QCD axion and axion-like particles are compelling candidates for galactic dark matter.
Theoretically, axions can convert into photons in the presence of a strong external magnetic field, which means it is possible to search for them experimentally.
One approach is to use radio telescopes with high-resolution spectrometers to look for axion-photon conversion in the magnetospheres of neutron stars.
In this paper, we describe the results obtained using a novel approach where we used the Green Bank Telescope (GBT) to search for radio transients produced by collisions between neutron stars and dark matter clumps known as axion miniclusters.
We used the VErsatile GBT Astronomical Spectrometer (VEGAS) and the X-band receiver (8 to 10~GHz) to observe the core of Andromeda.
Our measurements are sensitive to axions with masses between 33 and 42~$\mu$eV with $\Delta m_a = 3.8 \times 10^{-4}$~$\mu$eV.
This paper gives a description of the search method we developed, including observation and analysis strategies.
Given our analysis algorithm choices and the instrument sensitivity ($\sim$2~mJy in each spectral channel), we did not find any candidate signals greater than $5\sigma$.
We are currently implementing this search method in other spectral bands.
\end{abstract}

                              
\maketitle


\section{Introduction}


Astrophysical and cosmological observations suggest that approximately 27\% of the energy content of the universe resides in dark matter~\cite{Planck:2018vyg, Planck:2018lbu, ACT:2023kun, DES:2022urg}.
It seems dark matter seeded structure formation~\cite{eBOSS:2020yzd}.
Dark matter is responsible for the gravitational lensing of distant objects~\cite{Natarajan:2017sbo, Meneghetti:2020yif}.
There is also convincing evidence that it is driving the rotation curves of galaxies~\cite{Rubin:1980zd}.  
Despite its importance in the cosmological story, we do not know what dark matter is~\cite{Bertone:2004pz}.
If it is a new particle, then it can not belong to the Standard Model because it does not interact with light in the usual way.~\cite{Marsh:2024ury}

One candidate dark matter particle is the quantum chromodynamic (QCD) axion~\citep{Weinberg:1977ma, Wilczek:1977pj, Marsh:2015xka, DiLuzio:2020wdo}, a heretofore undetected particle predicted to exist by the Peccei-Quinn (PQ) mechanism, which solves the strong charge-parity (CP) problem in particle physics~\citep{Peccei:1977hh}.
Along with the QCD axion, other similar axion-like particles (ALPs) have also been proposed in theories that extend the Standard Model~\cite{Georgi:1986df, Arias:2012az}.
Although ALPs do not directly solve the Strong-CP problem, they share similar properties with the QCD axion such as their nature as a pseudo-scalar Goldstone boson.
QCD axions and ALPs are both compelling dark matter candidates because the particle lifetime exceeds the age of the universe, and on galactic or cosmological scales, they behave as a pressure-less non-interacting fluid~\cite{Preskill:1982cy, Abbott:1982af, Dine:1982ah}.
Throughout the paper, we will refer to QCD axions and ALPs generically as ``axions."

Theoretically, the axion has unavoidable feeble interactions with the Standard Model particles; the most notable being the axion-photon coupling, with a strength that is parameterized by a coupling constant $g_{a \gamma \gamma}$.
This coupling allows axions to convert into photons in the presence of an external magnetic field via the Primakoff-Raffelt effect~\cite{primakoff_1951,Raffelt:1987im}.
The photon production rate is proportional to the square of the magnetic field and the axion-photon coupling, so $P \propto ( g_{a \gamma \gamma} B )^2$, where $P$ is the power of the emerging photon signal and $B$ is the external magnetic field.
For non-relativistic axions (as is the case for axion dark matter), the energy of the photon generated in the aforementioned process is related to the axion mass $m_a$ via $E = m_a c^2 = h \nu $.
This implies $\nu = 0.242\,(m_a/\mu\text{eV})$~GHz.
The conversion process can be resonantly enhanced in the presence of an ambient plasma, which gives rise to an effective photon mass.
The conversion enhancement happens when this effective photon mass matches the axion mass energy~\cite{Raffelt:1987im}.

Ideal axion searches thus require (i) strong magnetic fields, (ii) large dark matter densities in order to generate a sizable photon flux, and (iii) an instrument capable of measuring both the intensity and the frequency of any emerging signal.
Laboratory-based experiments have demonstrated the potential to discover axions with haloscopes~\cite{Sikivie:1983ip, Sikivie:1985yu, Chadha-Day:2021szb, Semertzidis:2021rxs, Adams:2022pbo}.
Haloscopes are fantastic tools for axion characterization studies because the controlled experimental setup allows for precise measurement of both $m_a$ and $g_{a\gamma\gamma}$.
However, a given haloscope can typically probe a comparatively small portion of the large axion mass range that is theoretically allowed.
A complementary idea is to use radio telescopes to look for axion-photon conversion in the magnetosphere of a neutron star (NS)~\citep{Pshirkov:2007st, Safdi:2018oeu, Hook:2018iia}.
Available spectrometer backends with high frequency resolution can be used to probe a broad range of axion masses quickly.
In the neutron star magnetosphere, the conversion process is enhanced by the ambient plasma, and the signal strength is maximized by the enormous magnetic field ($\sim10^{10}$~T) which may be present~\citep{Pshirkov:2007st, Hook:2018iia, Safdi:2018oeu, Battye:2019aco, Leroy:2019ghm, Witte:2021arp, Battye:2021xvt}\footnote{See \citet{Lai:2006af} and \citet{Dessert:2022yqq} for the conversion in magnetized white dwarfs.}. 
The axions converted to photons in this way may belong to the smooth dark matter halo, or they may be sourced by the time-dependent electromagnetic fields of the neutron star itself~\cite{Prabhu:2021zve,Noordhuis:2022ljw,Noordhuis:2023wid,Khelashvili:2024sup}.

Some radio telescope-based axion searches have already been completed.
Small neutron star populations in the Milky Way have been observed in L-band (1 to 2~GHz) with the Green Bank Telescope (GBT) and in L-band and S-band (2 to 4~GHz) with the Effelsberg Radio Telescope~\citep{Foster:2020pgt}.
A single magnetar (PSR~J1745-2900) near the Galactic Center has been observed in C-band (4 to 8~GHz), X-band (8 to 12~GHz), and Ku-band (12 to 18~GHz)~\citep{Darling:2020uyo, Darling:2020plz, Battye:2021yue}.
\citet{zhou_2022} observed an isolated neutron star (J0806.4-4123) with MeerKAT from 769 to 1051~MHz, while \citet{Battye:2023oac} searched for time-dependent signals associated with the rotation of the pulsar J2144-3933 over a similar frequency range.
\citet{Foster:2022fxn} used archival C-band data from the GBT that was collected in a survey of the Galactic Center by the Breakthrough Listen project~\cite{merali_2015}.

An alternate but related detection strategy is to search for \emph{radio transients} caused by collisions between neutron stars and dark matter clumps known as axion miniclusters (AMCs)~\cite{Hogan:1988mp, Kolb:1993zz}.
AMCs are promising targets for indirect axion detection because of the increased dark matter density with respect to the galactic abundance, which enhances the radio power emitted when an AMC passes through a neutron star~\cite{Edwards:2020afl,Witte:2022cjj}.
Also, the unique time-variable nature of the signal could be used to differentiate the axion signal from other astrophysical signals and radio frequency interference (RFI) produced by satellites or other terrestrial transmitters.
AMCs are interesting theoretical targets, since they are predicted to form from QCD axion dark matter when the PQ phase transition happened during the radiation dominated epoch of the early Universe. 
This dark matter model is unique, since in principle the axion mass can be predicted from the relic density using cosmic string simulations~\cite{Hiramatsu:2012gg, Klaer:2017ond, Gorghetto:2018myk, Buschmann:2019icd, Gorghetto:2020qws, Buschmann:2021sdq, Hoof:2021jft, Saikawa:2024bta, Kim:2024wku}.

This paper describes the first AMC search, which was carried out in X band, corresponding to the axion mass range of 33 to 50~$\mu$eV.
The axion mass predicted from recent high resolution cosmic string simulations~\cite{Buschmann:2021sdq} is 40 to 180~$\mu$eV.
Since X band overlaps with this mass range, it is a very well motivated search path for axions.
In Sec.~\ref{sec:methods} we describe the anticipated signal and our observations.
We then present our results in Sec.~\ref{sec:results}; the final spectra are shown in Fig.~\ref{fig:final_spectra}. 
We discuss the implications of our findings and future work in Sec.~\ref{sec:discussion}. 
We conclude in Sec.~\ref{sec:conclusions}.


\section{Methods}
\label{sec:methods}


\subsection{Observational Overview}
\label{sec:observational_overview}


Detectable signals from AMC-NS collisions are likely to be rare because the high axion mass predicted (40 to 180~$\mu$eV) requires a very high plasma frequency to achieve resonance, which in turn requires very large magnetic fields in very young neutron stars.
To help alleviate this probability problem, we chose to observe the center of the Andromeda Galaxy (also called M31).
The precise number of neutron stars in M31 is not known, but a comparison with the Milky Way suggests a number $\mathcal{O}(10^9)$ might be a reasonable estimate~\cite{Sartore:2009wn}.
M31 is 180~arcmin across, and the telescope beam width is 1.4~arcmin (more in Sec.~\ref{sec:instrument}).
Assuming naively that the neutron stars are distributed evenly throughout the galaxy, we could have $\mathcal{O}(10^4)$ neutron stars in the telescope beam at all times.
It is likely that there are more neutron stars in the core of the galaxy where we chose to point, so this estimate is likely a lower limit.
Potentially, a non-zero fraction of the $\mathcal{O}(10^4)$ neutron stars in the telescope beam will be young with large magnetic fields and interacting with an AMC during our observation.
The penalty associated with observing M31 is the galaxy is 2.5~Mly (767 kpc) away and the signal is diluted as $1/d^2$. 
Here, $d$ is the distance between the observer and the event.
Nevertheless, we chose to take the large number of neutron stars per beam approach, thinking we need just one of them to produce a bright and detectable event.
More detail regarding our rationale is given in Sec.~\ref{sec:discussion}.

Spectroscopic observations made with radio telescopes could in principle lead to measurements of both $m_a$ and $g_{a\gamma\gamma}$.
However, a measurement of $g_{a\gamma\gamma}$ requires accurate neutron star and AMC population modeling, which is an active area of research (see, for example, \cite{Witte:2022cjj, DeMiguel:2023tlq}).
A measurement of the axion mass only requires a detection of the frequency of the emitted signal, which can be made with a high frequency resolution spectrometer.
Our study for now is focused only on searching for and measuring $m_a$.


\subsection{Approximate Signal Strength}
\label{sec:approximate_signal_strength}


Here, we present a signal strength calculation that is designed to assess detectability.
Therefore, this calculation should be taken as an order-of-magnitude estimate.
Note that the radio astronomy community commonly uses SI units.
We want to connect our signal prediction to our observations in the end, so our calculation was done using SI units rather than natural units.

We estimate the signal strength for axion-photon conversion from a single AMC-NS encounter using the Goldreich-Julian (GJ) model~\cite{Goldreich:1969sb}.
We consider a neutron star of radius $R_{\rm NS}$ and rotating with spin period $P$.
We assume that the magnetic dipole is aligned with the rotation axis of the neutron star (along the $z$-direction).
With these assumptions, the $r$ and $\theta$ components of the magnetic field are:
\begin{align}
   \label{eq:B_r}
    B_r(r,\theta) &= B_0 \left(\frac{R_{\rm NS}}{r}\right)^3 \cos \theta\,,\\
\label{eq:B_theta}
B_{\theta}(r,\theta) &= \frac{B_0}{2} \left(\frac{R_{\rm NS}}{r}\right)^3 \sin \theta\,,
\end{align}
and the $\phi$ component is zero~\cite{Witte:2021arp}.
Here, $B_0$ is the magnetic field strength at the north and south poles and $\theta$ is the polar angle measured from the rotation axis. 
The magnitude of the magnetic field is then:
\begin{equation}
    \label{eq:Bfield}
    B(r,\theta) = \frac{B_0}{4} \left(\frac{R_{\rm NS}}{r}\right)^3 \left(3\cos^2 \theta + 1\right)\,,
\end{equation}
while the $z$-component can be written:
\begin{equation}
    \label{eq:Bzfield}
    B_z(r,\theta) = \frac{B_0}{2} \left(\frac{R_{\rm NS}}{r}\right)^3 \left(3\cos^2 \theta - 1\right) \,.
\end{equation}
The neutron star is surrounded by a plasma with electron density close to the surface given by:
\begin{equation}
n_e(r, \theta) \approx \frac{4 \pi \epsilon_0}{e} \frac{\left|B_z(r,\theta)\right|}{P} \,,
\end{equation}
where $\epsilon_0$ is the vacuum permittivity, and $e$ is the electron charge.
In the Goldreich-Julian model, this expression for $n_e$ is axially symmetric, and it depends on the radial distance from the neutron star $r$ and the polar angle $\theta$.
Under these assumptions, the plasma frequency is:
\begin{equation}
    \label{eq:plasmafreq}
    \omega_p(r, \theta) = \sqrt{\frac{n_e(r,\theta) e^2}{m_e \epsilon_0}}\,,
\end{equation}
where $m_e$ is the electron mass~\cite{PhysRev.33.195}.
The radial and angular dependencies of $\omega_p$ derive from the dipole structure of the magnetic moment of the neutron star within the model.

Our desired simple estimate for the  luminosity from axion-photon conversion in the magnetosphere of a neutron star is~\cite{Pshirkov:2007st,Hook:2018iia}: 
\begin{align}
\begin{split}
\label{eq:power}
L_a \approx 2 \times c^2  \left[ g_{a\gamma\gamma} \, \frac{B_0/2}{\sqrt{\mu_0}} \left( \frac{{R_{\rm NS}}}{{R_c}} \right)^3 \right]^2 \\ \times \left[ \frac{2 \pi \hbar}{3 c} \frac{R_c}{m_a} \right] \times \left[ \, \rho_a(R_c){R_c}^2 v_c \, \right] \, ;
\end{split}
\end{align}
where $R_c$ is the radius at which the plasma frequency in the magnetosphere in Equation~\eqref{eq:plasmafreq} is equal to the axion mass ($\hbar \omega_p(R_c,\theta) = m_a c^2$); the velocity of an infalling axion at the conversion radius is
\begin{align}
\label{eq:velocity}
v_c \approx \sqrt{2 G M_\mathrm{NS}/R_c};
\end{align}
$\rho_a(R_c)$ is the axion density at the conversion radius;\footnote{Note here that the dimensionful coupling $g_{a\gamma\gamma}$ has units of $(E/L)^{-1/2}$.
In natural units, the coupling is typically reported in units of $E^{-1}$, for example $\mathrm{GeV}^{-1}$.
The quantity reported in this way is in fact $g_{a\gamma\gamma}/\sqrt{\hbar c}$ (with $\hbar = c = 1$).}
and we chose $\theta = \pi/2$ as an example polar angle.
The factor of 2 at the front is included because an infalling axion will cross the conversion surface twice, so the conversion probability doubles.
The expression in Equation~\eqref{eq:power} is only valid close to the resonance ($\hbar \omega_p(R_c,\theta) = m_a c^2$), where the mixing between the axion and photon is maximal.
Given the aforementioned assumptions, we obtain the following expression for the conversion radius, at which resonant conversion occurs:
\begin{align}
\label{eq:Rc}
R_c = R_{\rm NS} \left[ \frac{2 \pi \hbar^2 e}{m_e c^4} \frac{B_0}{m_a^2 P} \right]^{1/3}.
\end{align}
Non-resonant conversion ($R \neq R_c$) is expected to be sub-dominant in this case, as it receives an additional suppression due to the momentum mismatch between the axion and photon~\cite{Kelley:2017vaa,Sigl:2017sew}.

The scaling of Equation~\eqref{eq:power} can be understood intuitively.
The mass energy density of axions at the conversion radius is simply $\rho_a(R_c)$.
The rate at which axion mass energy crosses the conversion radius is then $\rho_a(R_c) \, R_c^2 \, v_c$.
The conversion probability scales as $\left[\,g_{a\gamma\gamma} B(R_c)\,\right]^2\times [R_c/m_a]$, where the first term determines the strength of the axion-photon interaction and the $R_c/m_a$ term is roughly $L_c^2$, where $L_c$ is the size of the region around $R_c$ in which conversion is possible.
If the plasma frequency varies rapidly with radius, then conversion is only possible in a very narrow region and the total conversion probability is low. More specifically, one finds that $L_c \sim \left|\,\mathrm{d}\omega_p/\mathrm{d}r\,\right|^{-1/2} \propto (R_c/m_a)^{1/2}$, where we have evaluated the derivative directly using Equation~\eqref{eq:Rc}, with $r = R_c$ and $\hbar \omega_p = m_a c^2$.
This term can also be understood in terms of the Compton wavelength of the axion $\lambda_a = 2 \pi \hbar/(m_a c)$~\cite{Pshirkov:2007st}, so that $L_c \sim (R_c \lambda_a)^{1/2} $. The larger the value of $m_a$, the smaller the wavelength of the axion field and the smaller the region over which the resonance condition can be achieved.
Finally, writing $B(R_c)^2 \propto [ \, B_0 (R_{\mathrm{NS}} / R_c )^3 \, ]^2$, we obtain the expression in Equation~\eqref{eq:power}.

Axion conversion is only possible when $R_c>R_{\rm NS}$.
In the Goldreich-Julian model, this implies a relationship between the $B$ field at the pole, the period, and the maximum axion mass for which there is a resonant signal.
For axions in X-band (33 to 50~$\mu$eV), resonance can comfortably be achieved for magnetic fields around $10^{14}\text{ G}$.
To obtain a rough scale for the radiated power, we assume $R_{\rm NS} = 10\text{ km}$ and $M_{\mathrm{NS}} = 1.4~M_{\odot}$.
For the axion density at large distances $\rho_a^\infty$ we take the mean AMC density $\rho_a^\infty = 7\times 10^{10}\text{ GeV cm}^{-3}$, corresponding to an overdensity $\delta=10$ at the time of formation~\cite{Kolb:1993zz},\footnote{The density is extremely large compared to typical dark matter halos like the Milky Way with $\rho\approx 0.4 \text{ (GeV/}c^2\text{) cm}^{-3}$, which is why AMCs are excellent sources.
The AMC mean density is roughly $\rho_a=140 \rho_{\rm eq}\delta^4$ where the matter density at equality is $\rho_{\rm eq}\sim \mathcal{O}(1\text{ eV}^4) \sim 5\times10^{4} \text{ (GeV/}c^2\text{) cm}^{-3}$.
This corresponds to a virialised object forming at equality when the overdensity is $\delta=1$, while if $\delta=10$ the AMC forms at $z\approx10z_{\rm eq}$~\cite{Kolb:1993zz,Ellis:2020gtq}.} which is comfortably dense enough for the AMCs to survive tidal stripping in the centre of M31~\cite{Kavanagh:2020gcy, Xiao_2021, DSouza:2024flu}, while not being too rare in AMC simulations~\cite{Ellis:2022grh}.
The axion density near the neutron star receives an additional enhancement due to gravitational focusing:
\begin{equation}
\label{eq:rho}
\rho_a(R_c) \approx \frac{v_c}{v_0}\rho_a^\infty  \sim \mathcal{O}(100) \,\rho_a^\infty\,,
\end{equation}
where $v_0$ is the axion velocity at large distances.
The virial velocity of axions inside an AMC is typically very small, so here we can take $v_0 \approx v_\mathrm{rel}$, the relative velocity of the AMC-NS encounter.
We further fix the coupling to be given by the QCD axion with anomaly ratio $E/N=0$, giving $g_{a\gamma\gamma}=1.6\times 10^{-14}\text{ GeV}^{-1}(40\,\mu\text{eV}/m_a)$.
Thus, inserting Eqs.~\eqref{eq:velocity}, \eqref{eq:Rc}, and \eqref{eq:rho} into Equation~\eqref{eq:power}, we obtain:
\begin{align}
\begin{split}
\label{eq:power_ref}
L_a &\approx L_0\left(\frac{g_{a\gamma\gamma}}{1.6 \times 10^{-14}\text{ GeV}^{-1}}\right)^2 \left(\frac{m_a}{40\,\mu\text{eV}}\right)^{5/3} \\
&\mkern-18mu\times \left(\frac{B_0}{10^{14}\text{ G}}\right)^{2/3} \left(\frac{P}{1{\rm\,s}}\right)^{4/3} \left(\frac{\delta}{10}\right)^4 \left(\frac{v_\mathrm{rel}}{220\,\mathrm{km/s}}\right)^{-1}\, ,
\end{split}
\end{align}
where $L_0 = 5\times 10^{19}\text{ W}$.
The luminosity given in Equation~\eqref{eq:power_ref} is the average luminosity expected assuming an AMC of uniform density.
The mean flux density assuming isotropic emission is:
\begin{equation}
\label{eq:flux}
S_\nu = \frac{1}{\Delta\nu} \frac{1}{ 4 \pi d^2} \, L_a\,,
\end{equation}
where $\Delta\nu$ is the bandwidth of the observation~\cite{Hook:2018iia}.
For M31, assuming $L_a = L_0$, $d = 770\,\mathrm{kpc}$, and $\Delta\nu = 92$ kHz (our frequency resolution, see Sec.~\ref{sec:instrument}), the signal from a single neutron star would be $S_{\nu} \approx 10^{-3}$~mJy, which is not detectable (more in Sec.~\ref{sec:instrument}).
However, if $\delta$ increases to $\delta = 100$, for example, or if $g_{a\gamma\gamma}$ increases to $10^{-12}\,\mathrm{GeV}^{-1}$, which could be possible for ALPs, then $S_{\nu}$ increases by a factor of $10^4$, and the signal becomes detectable.


\begin{figure}[t]
\centering
\includegraphics[width=\columnwidth]{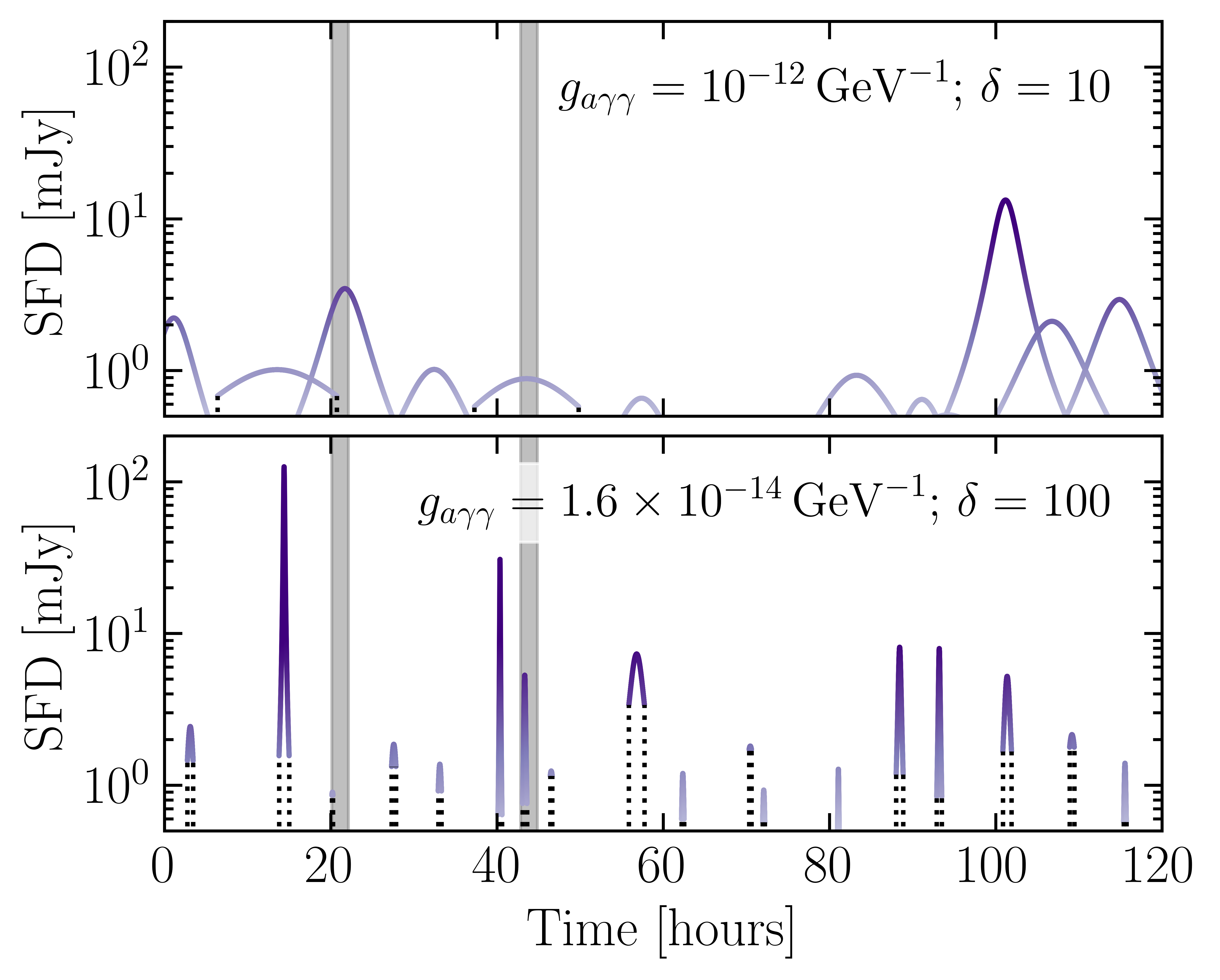}
\caption{
Calculated spectral flux densities (SFDs) from AMC-NS encounters shown for illustrative purposes.
Two example scenarios are considered.
In each panel, we assume a total of 20 encounters between neutron stars with polar B-fields $B_0 = 10^{14}\,\mathrm{G}$, rotation periods $P = 1\,\mathrm{s}$, and AMCs masses $M_\mathrm{AMC} = 10^{-10}\,M_\odot$.
We assume an axion mass of $m_a = 40\,\mu\mathrm{eV} = 9.7\,\mathrm{GHz}$.
In the upper panel, we assume AMCs with an overdensity parameter $\delta = 10$ and an axion-photon coupling $g_{a\gamma\gamma} = 10^{-12}\,\mathrm{GeV}^{-1}$, which is below the current CAST limits~\cite{CAST:2017uph}.
In the lower panel, we assume denser AMCs ($\delta = 100$) with coupling appropriate for the QCD axion ($g_{a\gamma\gamma} = 1.6\times 10^{-14}\,\mathrm{GeV}^{-1}$); the duration of these events is typically 30 minutes to one hour.
The vertical gray bands are representative of the duration and spacing of our observations.
Our $1\sigma$ sensitivity per spectral channel is approximately 2~mJy (see Sec.~\ref{sec:results}), so we should be able to observe signals like these.
}
\label{fig:AMC-NS_signals}
\end{figure}


For an AMC density profile $\rho \propto r^{-9/4}$~\cite{Ellis:2022grh}, the luminosity will vary with time as the neutron star transits the AMC and will be substantially enhanced for encounters passing close to the AMC center.
For illustrative purposes, in Fig.~\ref{fig:AMC-NS_signals}, we show how the spectral flux density changes as a function of time during an AMC-NS encounter.
In the upper panel, we assume an axion-photon coupling of $g_{a\gamma\gamma} = 10^{-12}\,\mathrm{GeV}^{-1}$, much larger than the value expected for the QCD axion and close to the current limits from the CAST experiment~\cite{CAST:2017uph}.
In the lower panel, we assume a smaller $g_{a\gamma\gamma} = 1.6 \times  10^{-14}\,\mathrm{GeV}^{-1}$ with an overdensity parameter of $\delta = 100$, leading to a larger dark matter density within the AMCs and a correspondingly larger flux.
These two example scenarios would give rise to a flux large enough to be detected in the current search, with a typical event timescale of hours to tens of hours.


\begin{figure}
\centering
\includegraphics[width=\columnwidth]{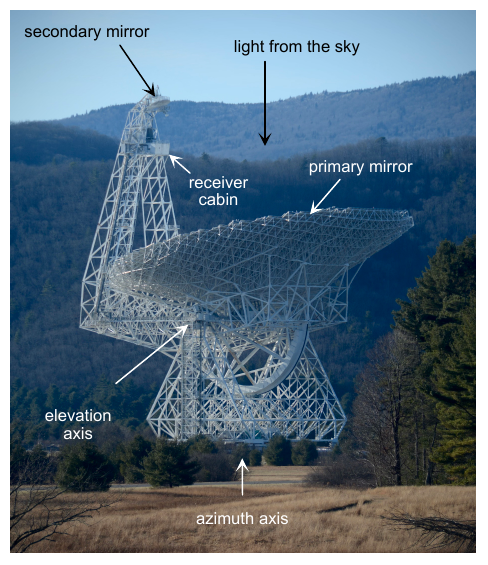}
\caption{
A photograph of the GBT pointed near zenith.
Light from the sky is focused into the X-band receiver in the receiver cabin with the parabolic primary mirror and the elliptical secondary mirror.
The projected diameter of the primary mirror is 100~m.
The telescope beam is pointed by moving the telescope in azimuth and elevation.
}
\label{fig:gbt}
\end{figure}


\begin{figure}
\centering
\includegraphics[width=\columnwidth]{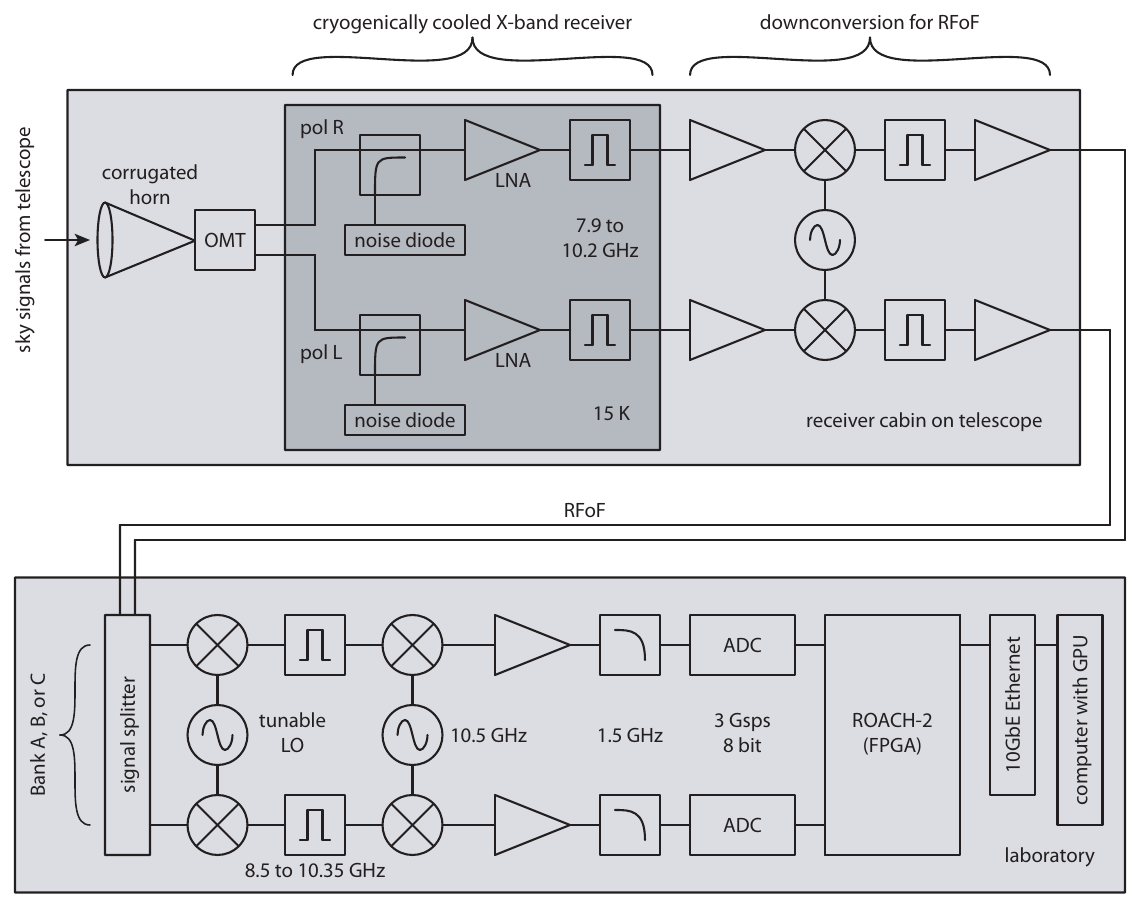}
\caption{
Schematic of the X-band receiver (top) and the VEGAS spectrometer (bottom).
The receiver is mounted in the receiver cabin at the Gregorian focus of the telescope (see Fig.~\ref{fig:gbt}).
The VEGAS spectrometer operates approximately 2~km away near the telescope Control Room.
Sky signals travel from the receiver to the spectrometer using RF over fiber (RFoF).
The X-band receiver is sensitive to right (R) and left (L) circular polarization.
More technical details are given in Sec.~\ref{sec:instrument}.
}
\label{fig:spectrometer}
\end{figure}


\subsection{Telescope, Receiver, and Spectrometer}
\label{sec:instrument}


We conducted our spectroscopic observations of M31 with the $100$~m Robert C.\ Byrd Green Bank Telescope at the Green Bank Observatory (GBO) in West Virginia (see Fig.~\ref{fig:gbt}).
For these observations we used the X-band receiver with the VErsatile GBT Astronomical Spectrometer (VEGAS) backend.
A schematic is shown in Fig.~\ref{fig:spectrometer}.
X band nominally spans from approximately 8.00 to 12.0~GHz.
The filter in the X-band receiver limits our measured spectra to a narrower frequency range from $7.90$ to $10.2$~GHz.
Therefore, the axion mass range for this study is 32.7 to 42.1~$\mu$eV.

VEGAS measures the spectrum of the sky by uniformly sampling the sky signal in the telescope beam at 3~Gsps with an analog-to-digital converter (ADC) and then computing the spectrum of this data with a polyphase filter bank~\cite{wolszczan_2021, jamot_2012}.
The polyphase filter bank ensures the response in each spectral channel is uniform.
Channel-response uniformity is important for spectral line observations because the observed line may not land at the center of the spectral channel.
VEGAS comprises eight dual-polarization digital spectrometer ``banks" that can be configured by the observer.
For our first observation, we used two of the available banks and centered them on $8.50$ and $9.60$\,GHz.
The spectral bank center is controlled with the tunable local oscillator (LO) shown in Fig.~\ref{fig:spectrometer}.
The other three observations used three banks with center frequencies $8.50$, $9.05$, and $9.60$\,GHz.
We decided to add a redundant, overlapping third bank between the original two banks because we wanted to be sure the band-pass filter shape (8.50 to 10.35~GHz) did not reduce sensitivity.
We chose to use VEGAS ``Mode $2$" for our observations.
In Mode $2$, each bank has 1,500~MHz of bandwidth and 16,384 spectral channels, which yields a spectral resolution of $92$~kHz.
The chosen spectral resolution sets the uncertainty on the axion mass at approximately $\Delta m_a = 3.8 \times 10^{-4}$~$\mu$eV.

We used the GBT sensitivity calculator to forecast the performance of the instrument for our observations~\cite{sensitivity_claculator}.
This calculator uses a modified version of the classic radiometer equation~\cite{maddalena_2022}.
Given our chosen spectrometer setup and our observing time award (see Sec.~\ref{sec:observation_details}), the $1\sigma$ error per spectral channel should be approximately 1.8~mJy.
Therefore, for a $5\sigma$ detection, we would need an AMC-NS collision signal greater than 9.0~mJy, which means this particular X-band search is limited to larger values of $g_{a\gamma\gamma}$ and/or $\delta$ (see Fig.~\ref{fig:AMC-NS_signals}).
The forecasted sensitivity is consistent with the computed error in our measurements (more in Sec.~\ref{sec:results}).


\subsection{Observation Details}
\label{sec:observation_details}


The anticipated duration of the AMC-NS collision signal we are searching for is between hours and days (see example signal in Fig.~\ref{fig:AMC-NS_signals}).
It is not possible to continuously observe M31 for days at a time from the GBO, so we chose to make a series of two-hour observations spread out over several months (see Table~\ref{tab:obs_details} for the exact dates).
This approach allowed us to probe multiple timescales.

Our observation method consisted of ``on-off" pointings.
We first collected data while pointing at and tracking a target (the ``on" pointing).
Then, we slewed the telescope so the target was outside the telescope beam and made similar observations (the ``off" pointing).
Data from the on-target pointing includes signals from atmospheric emission, the interstellar medium (ISM), the intergalactic medium (IGM), the target, and background signals from the cosmic microwave background (CMB), for example.
Data from the off-target pointing ideally have all of the same signals because we only moved the telescope beam approximately 1~deg, but they do not have signal from the target itself.
Therefore, with this approach, it is possible to reveal the spectrum of the target alone, which is the signal we want to measure, by subtracting off-target data from on-target data (more in Sec.~\ref{sec:baseline_alignment}).


\begin{figure}
\centering
\includegraphics[width=\columnwidth]{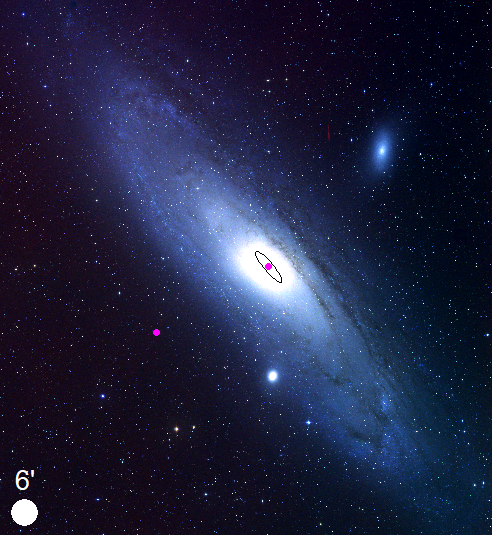}
\caption{
Hubble Space Telescope image of M31.
The two magenta dots mark our on- and off-target pointings (see Table~\ref{tab:targets} for precise coordinates).
The diameter of each dot represents the telescope beam size, which has a 1.4~arcmin FWHM. 
The black ellipse was added as a distance scale.
It is actually a circle with a radius of~1 kpc but appears as an ellipse in the image because of the inclination angle of M31.
With this study, we are simultaneously observing all of the neutron stars in the telescope beam, which we estimate could be $\mathcal{O}(10^4)$.
}
\label{fig:andromeda}
\end{figure}


\setlength{\tabcolsep}{5pt}
\renewcommand{\arraystretch}{1.2}
\begin{table}[h]
\centering
\begin{tabular}{c|c|c}
Target       & RA [\,hms\,] & Dec [\,dms\,] \\
\hline
on M31       & 00:42:44.330 & +41:16:07 \\
off M31      & 00:45:00.000 & +41:02:00 \\
on $3$C$48$  & 01:37:41.300 & +33:09:35.080 \\
off $3$C$48$ & 01:33:57.679 & +29:14:23.386 \\
\end{tabular}
\caption{
Celestial coordinates of the targets we used in this study.
All positions are J2000 coordinates.
The on- and off-M31 pointings are shown as the magenta dots in Fig.~\ref{fig:andromeda}.
Our off-target pointing for $3$C$48$ was specifically chosen to be an offset from $3$C$48$ in Galactic coordinates of $\delta l = 0.0$ and $\delta b = -4.0$.
This shift works out to the right ascension and declination coordinates reported above.
}
\label{tab:targets}
\end{table}


For each of our observations, we made on-off pointings of two targets: a known radio source for calibration ($3$C$48$) and the center of M31 (see Fig.~\ref{fig:andromeda}).
Table~\ref{tab:targets} gives the precise pointing coordinates.
$3$C$48$ was observed for approximately $15$ minutes during each two-hour observing session with a $30$ second on-target, $30$ second off-target cadence.
The remainder of the observing time was used for setting up the instrument, slewing the telescope, and observing M31 with a $60$ second on-target, $30$ second off-target cadence.
We achieved an average observing efficiency of 38\%.
%
%
The telescope elevation angle range for each observation is given in Table~\ref{tab:obs_details}.

A series of two-second integrations was recorded during each pointing.
For additional calibration, signal from a noise diode was added to the sky signal in the receiver using a directional coupler (see Fig.~\ref{fig:spectrometer}).
This diode signal was switched on and off with a period of two seconds during observations.
In each two-second integration, $\mathcal{O}(10^5)$ spectra were averaged together inside the spectrometer hardware, yielding a single measured spectrum for that integration.
%
%
As an example, a 60~sec pointing on M31 consists of 30 integrations; 15 of these integrations have the diode on.
Each observation consists of approximately 20 to 30 of these pointings.
The precise number of integrations for each observation is given in Table~\ref{tab:obs_details}.

The data collected during Observation~3 was excessively and unusually noisy, so we chose to abandon it.
The data is severely corrupted by RFI. 
The elevation angle for Observation 3 is significantly lower than the others, thus, the data from this observation is more corrupted by Earth's atmosphere.


\setlength{\tabcolsep}{5pt}
\renewcommand{\arraystretch}{1.6}
\begin{table*}[t]
\centering
\begin{tabular}{c|c|c|c|c|c|c|c}
            & Date      & Time      & Opacity    & Scans  & Integrations & System           & Elevation \\ [-0.06in]
Observation & [\,UTC\,] & [\,UTC\,] & at 10\,GHz & (M31)  & (on M31)     &  Temperature [K] &  Min/Max [Deg] \\ 
\hline
1 & 2022-02-04 & 20:08:34.00 & 0.017 & 23 & 690  & $30_{-0.8}^{+2.0}$ & $81/87$\\ 
2 & 2022-02-05 & 18:50:51.00 & 0.011 & 29 & 870  & $28_{-1.1}^{+1.0}$ & $71/83$\\ 
3 & 2022-04-01 & 12:12:25.00 & 0.012 & 38 & 1140 & $34_{-5.3}^{+6.6}$ & $37/51$\\ 
4 & 2022-07-10 & 10:26:12.00 & 0.041 & 18 & 540  & $31_{-1.3}^{+1.8}$ & $82/87$\\ 
\end{tabular}
\caption{
Observation details.
Each of the four observations started at the time in the table and lasted two hours.
During these observations, we executed the observing plan described in Sec.~\ref{sec:observation_details}.
Note that during Observation~\#4 there was moderate rainfall, which increased the atmospheric opacity.
The ``System Temperature" column gives the minimum and maximum values over course of each observation, and it is not a measurement uncertainty.
}
\label{tab:obs_details}
\end{table*}


\begin{figure}
\centering
\includegraphics[width=\columnwidth]{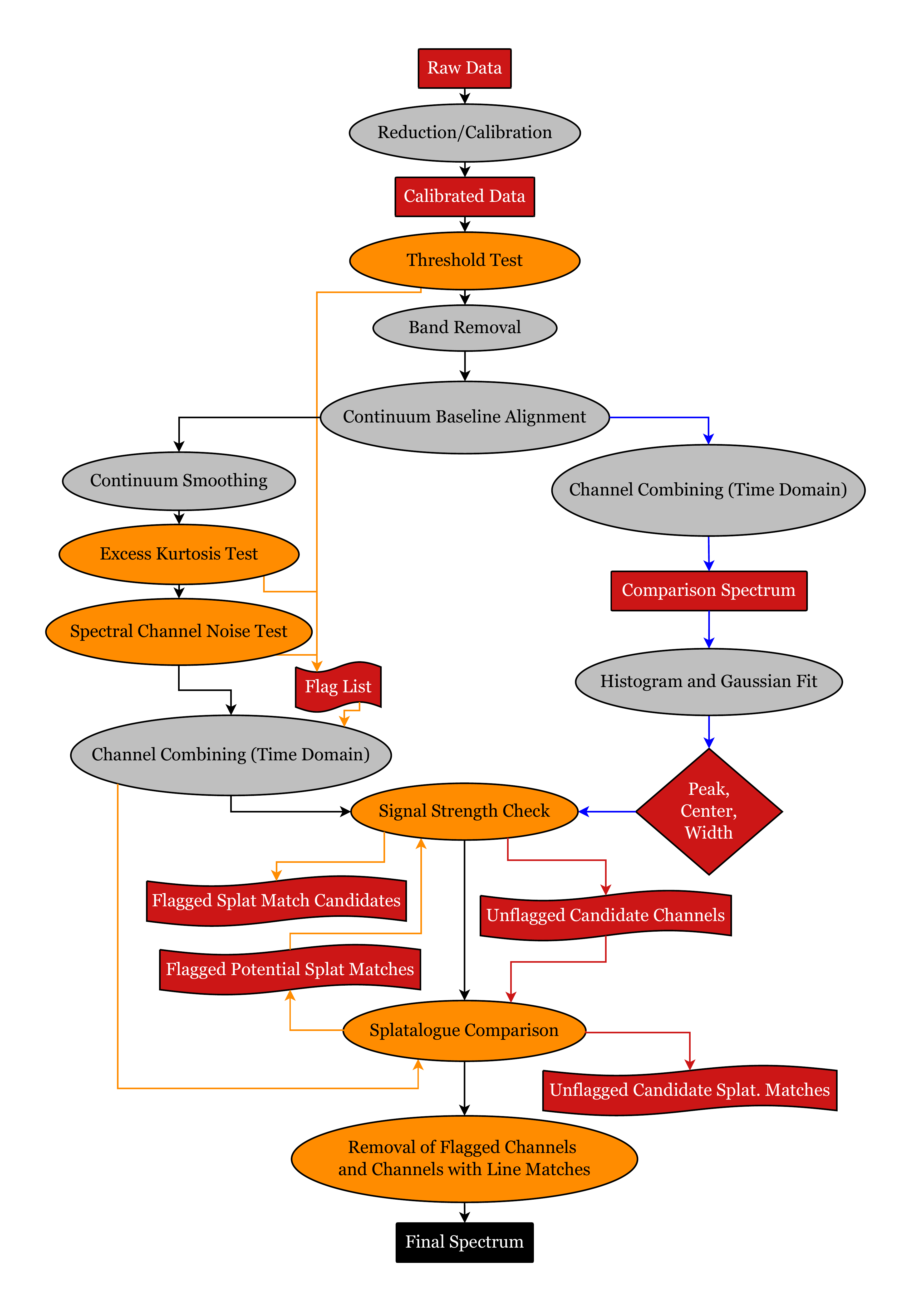}
\caption{
A flow chart that describes our analysis pipeline.
The grey ellipses represent a modification to the data.
The orange ellipses represent a flagging operation.
The red rectangles represent modified data that is used in other steps in the pipeline.
The red flag boxes represent lists of flagged channels.
The black arrows show the path of the data as it goes through the pipeline.
The orange arrows show the path of flagged channels.
The blue arrows show the path of the data used for assessing signal validity.
The red arrows indicate the path of spectral channels with a signal $3\sigma$ beyond the center of a Gaussian fit to a histogram of the median spectrum after the ``Continuum Baseline Alignment Operation."
See Sec.~\ref{sec:results} for more detail.
}
\label{fig:flowchart}
\end{figure}


\section{Results}
\label{sec:results}


\subsection{Calibration}
\label{sec:calibration}


Each observation is organized into ``scans" that consist of two telescope pointings, one on target and one off target.
We chose to use the noise diode for our calibration, so hereafter, only the M31 scans are considered.  
The data from each pointing consist of a series of the aforementioned two-second integrations.
In each scan, the on-target pointing yields 30 integrations, and the off-target pointing yields 15 integrations. 
We always performed the on-target pointing first followed by the off-target pointing.
This pattern continued for the entire time we observed our target. 

The data for each of our four observations were stored in Single Dish FITS (SDFITS) files, which is a conventional radio astronomy file format. 
Each SDFITS file contains all of the integrations for one spectrometer bank from one observation.
To analyze our data, we first used a custom Python function to parse the SDFITS files and sort the data into seven arrays.
One of these arrays contains the center frequencies of each spectral channel $f_\nu$.
There is a diode temperature array $T^\mathrm{diode}_{\nu,s,i}$ that gives the brightness temperature of the calibration diode for each spectral channel in each integration.
Our average diode brightness temperature was $1.8$~K. 
The $\nu$ index indicates the spectral channel, $s$ indicates the scan number, and $i$ indicates the integration number.
The average diode temperature for each integration in each scan is $\bar{T}^\mathrm{diode}_{s,i} = \langle T^\mathrm{diode}_{\nu,s,i} \rangle_\nu$.
We use $T^\mathrm{cal} = \langle \bar{T}^\mathrm{diode}_{s,i} \rangle_{s,i}$ in our calibration process for each bank in each observation. 
There is a telescope elevation angle array $\epsilon_{s,i}$.
Lastly, there are four source arrays: $M_{\nu,s,i}^{+}$ $M_{\nu,s,i}^{-}$, $O_{\nu,s,i}^{+}$, $O_{\nu,s,i}^{-}$. 
The $O_{\nu,s,i}^{+}$ array is the data collected when the telescope is pointed at the ``off M31" position with the noise diode switched on (indicated with the $+$ sign); $O_{\nu,s,i}^{-}$ is the same but with the noise diode switched off (indicated with the $-$ sign).
The $M_{\nu,s,i}^{+}$ array is the data collected when the telescope is pointed at the ``on M31" position with the noise diode switched on; $M_{\nu,s,i}^{-}$ is the same but with the noise diode switched off.
The four source arrays are sums of the R and L polarizations (see Fig.~\ref{fig:spectrometer}). 

Our calibration function follows the method described in~\citet{GBTIDL} and takes the aforementioned seven arrays as inputs\footnote{Originally, we planned on using a calibration method similar to the one described in~\citet{Abitbol:2018}, which is why we also observed 3C48 during each observation. }.
First, the function computes the system temperature for each scan $T^\mathrm{sys}_{s}$ using the equation
%
%
\begin{equation}
\label{eq:tsys}
    T^\mathrm{sys}_{s} = T^\mathrm{cal}\times\frac{\mathrm{med}(O_{\nu,s}^{-})_\nu}{\mathrm{med}(O_{\nu,s}^{+} - O_{\nu,s}^{-})_\nu} + \frac{T^\mathrm{cal}}{2},
\end{equation}  
where $O_{\nu,s}^{+}$ is the median spectrum for all integrations in scan $s$ with the noise diode on
\begin{align}
\label{eq:O_plus}
O_{\nu,s}^{+} = \mathrm{med}(O_{\nu,s,i}^{+})_i,
\end{align}
and $O_{\nu,s}^{-}$ is the median spectrum for all integrations in scan $s$ with the noise diode off
\begin{align}
\label{eq:O_minus}
O_{\nu,s}^{-} = \mathrm{med}(O_{\nu,s,i}^{-})_i.
\end{align}
The antenna temperature $T^a_{\nu,s,i}$ is then computed for each spectral channel in each integration across all scans with
\begin{align}
\label{eq:ta}
    T^a_{\nu,s,i} = T^\mathrm{sys}_{s}\times\frac{M_{\nu,s,i}-O_{\nu,s}}{O_{\nu,s}},
\end{align}
where
\begin{align}
\label{eq:sig}
    M_{\nu,s,i} = \frac{1}{2}(M_{\nu,s,i}^{+} + M_{\nu,s,i}^{-}),
\end{align}
and
\begin{align}
\label{eq:ref}
O_{\nu,s} = \frac{1}{2} \left[ \, O_{\nu,s}^{+} + O_{\nu,s}^{-} \, \right].
\end{align}
We calibrated all of the on-target integrations in a scan $M_{\nu,s,i}$ with the median of the adjacent off-target integrations $O_{\nu,s}$.
Finally, the data were converted to units of janskys with 
\begin{align}
\label{eq:jansky_convert}
    S_{\nu,s,i} = \frac{T^a_{\nu,s,i}}{2.85}\times\frac{e^{\tau_0/\mathrm{sin}(\epsilon_{s,i})}}{\eta_a\times\eta_l},
\end{align}
where $\eta_l = 0.99$ is the correction for rear spillover, ohmic loss, and blockage efficiency, and $\eta_a = 0.71$ is the aperture efficiency~\cite{observers_guide}. 
We used the aperture efficiency of the GBT at 10~GHz. 
The measured values for the atmospheric opacity at the zenith $\tau_0$ for each observation are listed in Table~\ref{tab:obs_details}.
Each integration in a scan has a corresponding elevation angle $\epsilon_{s,i}$, so each integration is individually converted from units of kelvin to janskys. 


\begin{figure*}
\centering
\includegraphics[width=0.95\textwidth]{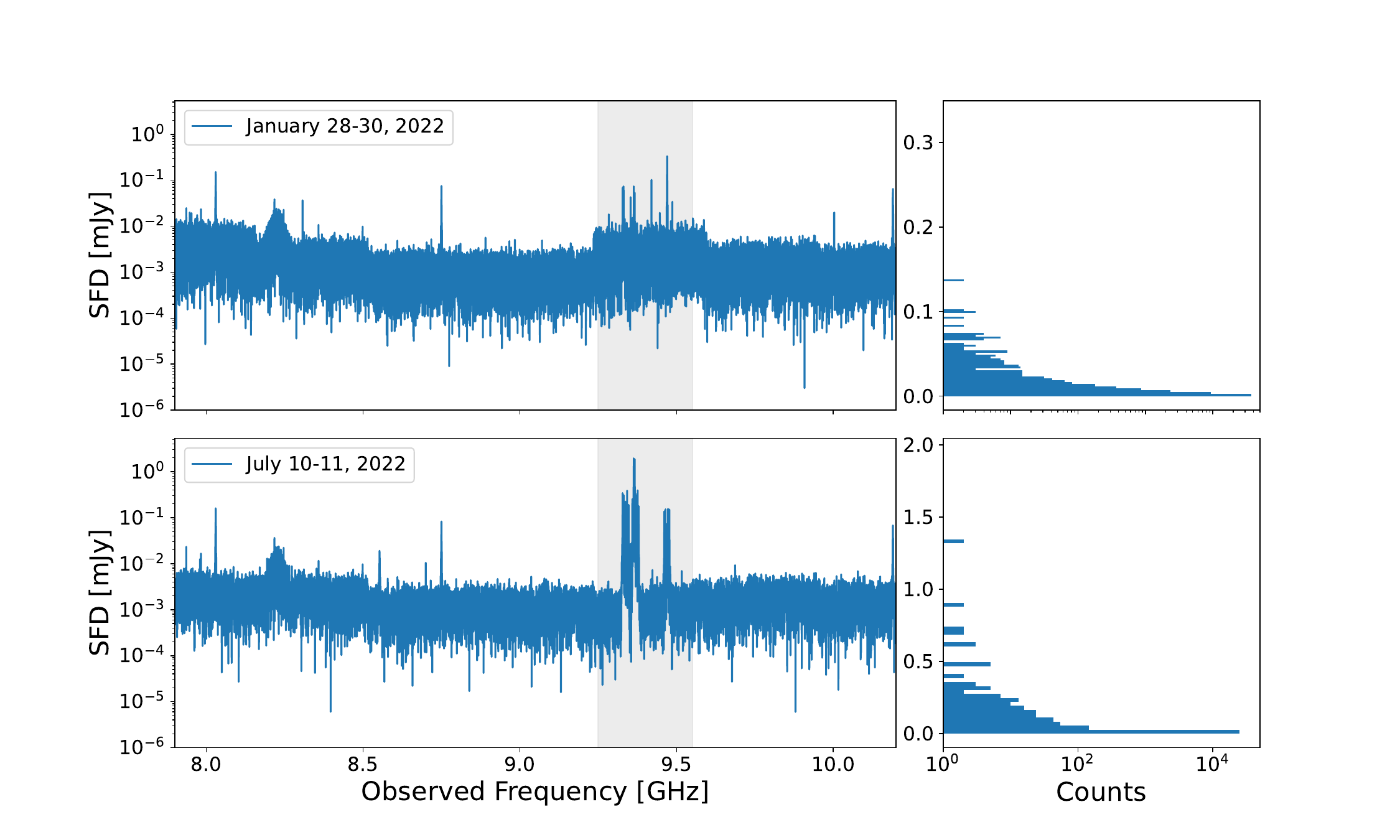}
\caption{
Time-averaged X-band spectra measured by GBO with an omnidirectional antenna mounted near the GBT.
Bright features in these spectra are probably from RFI, so these spectra were useful for helping us identify spectral channels in our data that should be removed from the axion search.
Data from the omnidirectional antenna were not collected synchronously with our observations (more in Sec.~\ref{sec:future_experiment}). 
The spectra shown here were collected close to our observation dates, however, (see Table~\ref{tab:obs_details}) on January 28-30, 2022 (UTC) (top panel) and July 10-11, 2022 (UTC) (bottom panel).
Note that the electromagnetic spectrum between 9.3 and 9.5~GHz (vertical gray bar) is allocated for Earth exploration satellites, radiolocation, radionavigation, and meteorology.
We chose to remove this spectral band from our data (see Sec.~\ref{sec:band_removal}).
Histograms of these spectra are shown on the right. 
}
\label{fig:RFI}
\end{figure*}


\subsection{Data Analysis Pipeline}
\label{sec:data_analysis}

Our data analysis pipeline searches for viable channels in the measured spectra that might contain signal from an AMC-NS collision, and it disqualifies problematic spectral channels.
Given the signal model in Equations~\ref{eq:power}~\&~\ref{eq:power_ref} and the forecasted AMC-NS collision signal plotted versus time in Fig.~\ref{fig:AMC-NS_signals}, we anticipated that the signal we are searching for in our measured spectra should be narrow; have a constant or slowly varying amplitude over each two-hour observation; and the noise associated with the signal should be Gaussian.
Our data analysis pipeline is focused on searching for signals that match this description.
The sensitivity of each observation is ideally limited by noise in each spectral channel coming from photon noise and noise from the croygenically cooled low-noise amplifier (LNA).
Spectral channels were disqualified if they appeared to contain RFI, if they were associated with known molecular or atomic emission lines, or if the spectral channel was corrupted by instrumental effects (from gain variations or intermodulation products, for example).
Broad spectral features that did not look like candidate AMC-NS collision signals were subtracted.
The data analysis pipeline we wrote is visually represented in the flow chart shown in Fig.~\ref{fig:flowchart}.
The individual pipeline operations are described below.
It is important to note that signal from an AMC-NS collision could exist in disqualified spectral channels, but we can not credibly test the AMC-NS collision hypothesis with this data set.


\subsubsection{Threshold Test}
\label{sec:threshold_test}

Any signal that originates from axion-photon conversion in M31 is not expected to be much brighter than approximately 100~mJy. 
Therefore, we disqualified spectral channels in individual integrations that contained signals significantly brighter than 100~mJy because we did not want these bright signals to bias downstream fitting algorithms or encumber downstream histogram operations.
We chose a conservative 10~Jy cutoff for these unwanted bright signals, which we suspect were produced by RFI.
Note that this ``Threshold Test'' does not disqualify entire spectral channels.
Instead, it removes problematic data points from the analysis.
Spectral channels flagged by this test were added to the ``Flag List'' (see Fig.~\ref{fig:flowchart}).


\subsubsection{Band Removal Operation}
\label{sec:band_removal}

A spectral region between 9.25 and 9.55~GHz consistently contained bright signals throughout all of our observations.
These signals can be seen in the spectra plotted in Fig.~\ref{fig:RFI}, for example.
The Federal Communications Commission (FCC) maintains a frequency allocation table that lists spectral bands designated for communication~\cite{FCC_2015}.
The table covers frequencies between 9~kHz and 275~GHz.
We used this table to help us identify and understand the detected signals between 9.25 and 9.55~GHz.
The FCC Frequency Allocation Table shows the electromagnetic spectrum between 9.3 and 9.5~GHz is allocated for Earth exploration satellites, radiolocation, radionavigation, and meteorology.
Therefore, we suspect the strong signals we observed in this band were produced by RFI.
Consequently, this ``Band Removal Operation" disqualified all spectral channels between 9.25 and 9.55~GHz in all observations (see Fig.~\ref{fig:flowchart}).


\subsubsection{Continuum Baseline Alignment Operation}
\label{sec:baseline_alignment}

The on-target data minus the off-target data should reveal the spectrum of the target alone, as discussed in Sec.~\ref{sec:observation_details}.
However, we found the difference spectra often contained residual continuum signal, probably from atmospheric emission variations produced by telescope elevation changes that happened between the on-target and the off-target pointings, and/or continuum slope differences introduced by readout electronic instabilities.
Since we are looking for narrow spectral lines and not broad continuum features, these continuum residuals were subtracted.
To remove the residuals, we fit a line to each integration yielding two fit parameters: an offset and a gradient.
We then constructed a residual continuum model from these parameters and subtracted it from each integration. 
The linear fit was done after the ``Threshold Test" and the ``Band Removal Operation" because we did not want the strong signals identified by these tests to bias the fitting algorithm.
It is important to note that this ``Continuum Baseline Alignment Operation" removes both the unwanted residuals and the continuum spectrum for M31, but it should preserve any narrow spectral features, which are the target of this study.
The primary consequence of this operation is the spectral baseline for each scan has a mean equal to zero. 


\subsubsection{Continuum Smoothing Operation}
\label{sec:continuum_smoothing}

After the linear continuum model was subtracted, some spectra still contained broad spectral features that were biasing the ``Spectral Channel Noise Test" that will be discussed below in Sec.~\ref{sec:spectral_channel_noise_test}.
The width of these features was greater than 10 spectral channels, but much less than 16,384 (see Sec.~\ref{sec:instrument}).
To remove these features we ran a customized boxcar smoothing operation designed to flatten the continuum without biasing narrow-band signals.
Here, narrow band means approximately one to three spectral channels wide.
The boxcar smoothing operation was performed by first computing a median spectrum for each observation, using all scans and integrations. 
From the median spectra, we computed a template spectrum that contained the ``rolling'' median values in a window 10 spectral channels wide. 
Each value in this template spectrum is the median of the 10 closest spectral channels. 
Therefore, the resulting template spectrum is a smooth version of the median spectrum, which we then subtracted from each integration. 
This operation resulted in final spectra without large-scale spectral ``wiggles.'' 
The 10-channel window size for the boxcar was chosen because it was the smallest window size that does not remove signals we are looking to detect. 
For example, the smoothing operation did not bias what appears to be a H(131)$\gamma$ feature we detected, which will be discussed later in Sec.~\ref{sec:spectral_feature_examples}.
This apparent H(131)$\gamma$ feature looks similar to the axion signals we are trying to find, so it served as a touchstone when developing this ``Continuum Smoothing Operation."


\begin{figure*}
\centering
\includegraphics[width=\textwidth]{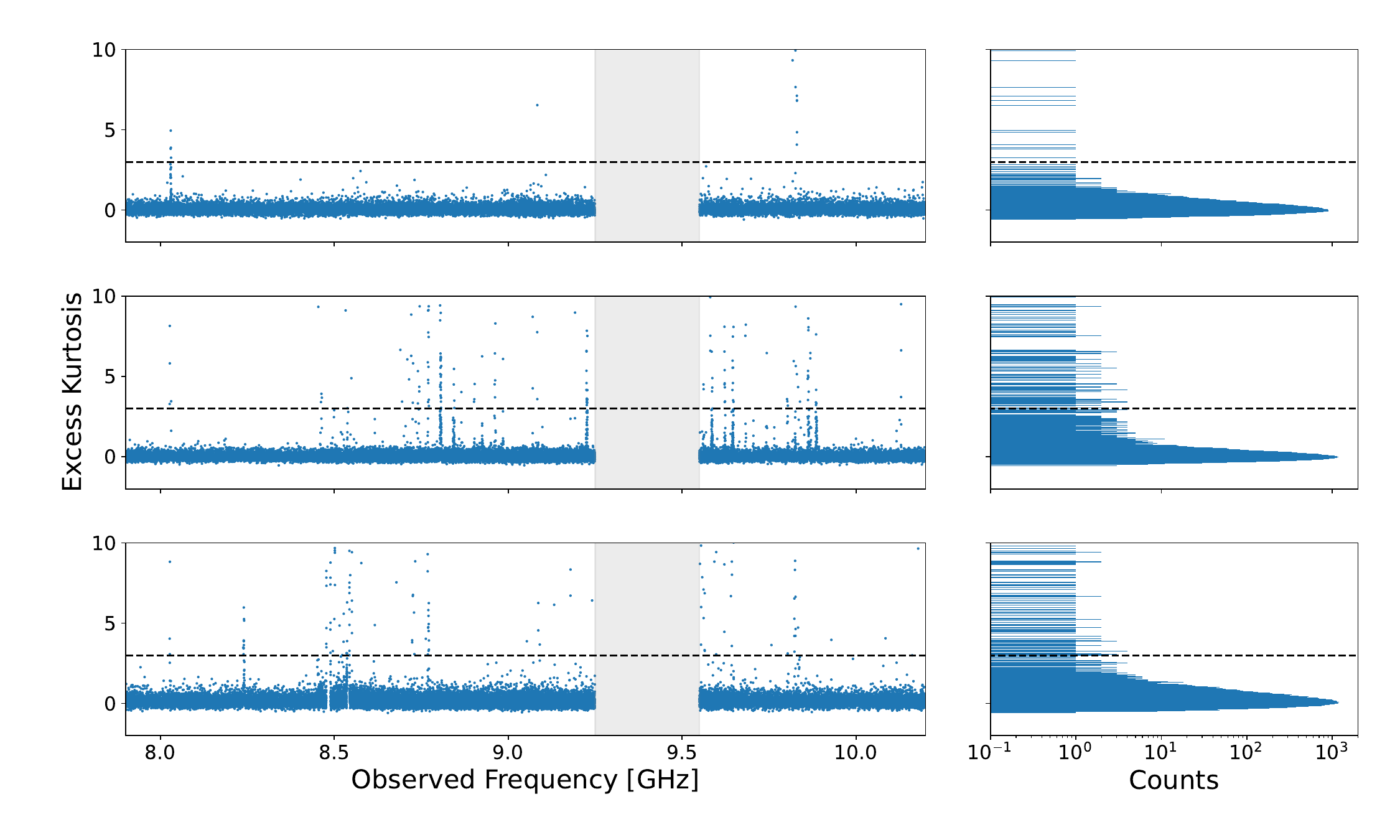}
\caption{
A graphical representation of the result from the ``Excess Kurtosis Test" for our observations (see Sec.~\ref{sec:excess_kurtosis_test}). 
The top, middle, and bottom panels are the first, second, and fourth observations, respectively.
The blue points show the excess kurtosis computed from the time-domain signal in each channel. 
The dashed, black line is our chosen threshold (excess kurtosis = 3). 
A histogram of each excess kurtosis spectrum is shown in the right column.
Note that the vast majority of points in the spectra have Gaussian noise (excess kurtosis $\approx$ 0), as expected.
}
\label{fig:kurtosis}
\end{figure*}


\subsubsection{Excess Kurtosis Test}
\label{sec:excess_kurtosis_test}

The ``Excess Kurtosis Test" is our primary method for identifying signals in our data that were produced by RFI.
It is based on the spectral kurtosis estimator developed by~\citet{nita_2010}.
Members of our team successfully used this spectral kurtosis technique in previous C-band observations~\cite{Abitbol:2018}.
For this test, the kurtosis was computed using all the viable data in a given spectral channel (all scans, all integrations).
If the data in the spectral channel has Gaussian noise, then the computed excess kurtosis should be zero.
We chose to disqualify spectral channels with excess kurtosis greater than three. 
Essentially, this test identifies channels that have signals with a non-Gaussian distribution. 
Spectral channels flagged by this test were added to the ``Flag List'' (see Fig.~\ref{fig:flowchart}).
The result from this test is shown in Fig.~\ref{fig:kurtosis}.


\begin{figure*}
\centering
\includegraphics[width=\textwidth]{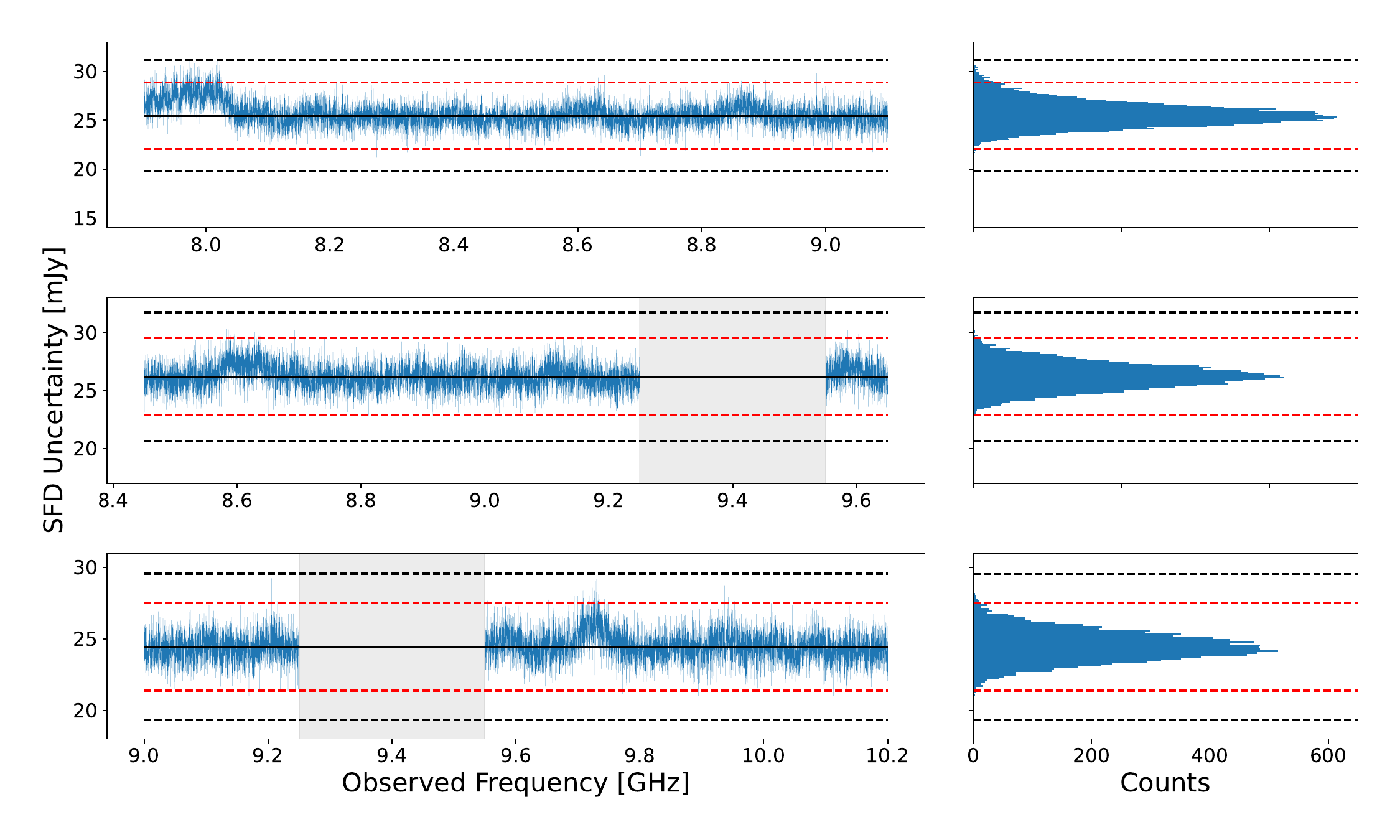}
\caption{
A graphical representation of the ``Spectral Channel Noise Test" described in Sec.~\ref{sec:spectral_channel_noise_test}. 
The data plotted here come from Observation~2 (spectral banks A, B, and C from top to bottom).
The blue points in the left panels are the widths of a Gaussian (1$\sigma_{\nu}$) estimated from a fit to a histogram of the time-domain data in each spectral channel. 
The right panels show histograms of these 1$\sigma_{\nu}$ values. 
The dashed red and black lines correspond to the $3\sigma_t$ and $5\sigma_t$ thresholds we used for spectral channel flagging, which were computed from Gaussian fits to the histograms on the right.
Any channel with $1\sigma_{\nu}$ values beyond these chosen thresholds is flagged, resulting in separate $3\sigma_t$ and $5\sigma_t$ flag lists. 
The solid, black line is the center of the Gaussian fit $\bar{\sigma}_{\nu}$, which shows the overall noise level for the continuum.
The downward line in center of each spectrum is a known instrumental artifact caused by the architecture of the ADC in the VEGAS spectrometer~\cite{observers_guide}.
}
\label{fig:gausscut}
\end{figure*}


\subsubsection{Spectral Channel Noise Test}
\label{sec:spectral_channel_noise_test}

If a spectrum contains only continuum emission, then the noise level in each spectral channel should be approximately the same.
Noise in a spectral channel that is larger than expected could indicate the presence of a non-continuum signal. 
To estimate the noise in each spectral channel, a histogram of all the of the data for that singular channel (all scans and integrations) was made using Freedman–Diaconis binning. 
This binning choice ensures the histogram has sufficient resolution along the abscissa.
Then, a one-dimensional Gaussian model was fit to this histogram with a least squares fitting algorithm.
The fit returned the peak, center, and width ($\sigma_{\nu}$) of the best-fit Gaussian. 
The center parameter gives an estimate of the mode for that spectral channel, and the width parameter gives the noise level for that channel.
The histogram and Gaussian fit were computed for each channel in each observation. 
We chose to use the histogram and Gaussian-fit approach rather than computing the spectral channel variance because we found this approach is less sensitive to bias from outliers.

To determine the flagging threshold, a histogram of all of the $1\sigma_{\nu}$ widths for all of the channels in a data set was computed using the same binning technique. 
A one-dimensional Gaussian model was then fit to this histogram, yielding the peak, center, and width ($\sigma_t$) of this uncertainty spectrum as a whole. 
Here, the center parameter gives the average $\sigma_{\nu}$ for the spectrum $\bar{\sigma}_{\nu}$, and $\sigma_t$ provides information about how spectral channel noise varies across the spectra.
The $\bar{\sigma}_{\nu}$ values we found for Observation 1, 2, and 3 are given in Table~\ref{tab:sigma_nu_bar}.
The thresholds we used for spectral-channel flagging were $3\sigma_t$ and $5\sigma_t$.
Any spectral channel with $\sigma_{\nu}$ values beyond these chosen thresholds was flagged, resulting in separate $3\sigma_t$ and $5\sigma_t$ flag lists (see Fig.~\ref{fig:flowchart}).  
Any spectral channel with $\sigma_{\nu}$ values beyond these chosen thresholds was flagged (see Fig.~\ref{fig:flowchart}). 
An example output from the ``Spectral Channel Noise Test" is shown in Fig.~\ref{fig:gausscut}. 


\setlength{\tabcolsep}{5pt}
\renewcommand{\arraystretch}{1.2}
\begin{table}[h]
\centering
\begin{tabular}{c|c|c|c|c}
& & Bank A & Bank B & Bank C \\
\hline
& Observation 1              & 27     & 29     & N/A \\
$\bar{\sigma}_{\nu}$ [\,mJy\,] & Observation 2              & 25     & 26     & 24 \\
& Observation 4              & 27     & 29     & 27 \\
\hline
& Observation 1              & 1.49      & 1.41      & N/A \\
$\sigma_{t}$ [\,mJy\,]& Observation 2              & 1.14      & 1.11      & 1.02 \\
& Observation 4              & 1.58      & 1.67      & 1.56 \\
\hline
\end{tabular}
\caption{
Noise estimates computed for the ``Spectral Channel Noise Test."
Here, $\bar{\sigma}_{\nu}$ is the average spectral channel noise in each spectrometer for each observation, and $\sigma_t$ is used for defining the threshold.
See Sec.~\ref{sec:spectral_channel_noise_test} and Fig.~\ref{fig:gausscut} for more detail.
}
\label{tab:sigma_nu_bar}
\end{table}


\subsubsection{Splatalogue Test}
\label{sec:splatalogue_test}

We cross-correlated our observations with the NRAO Spectral Line Catalog (Splatalogue)~\cite{splatalogue}.
Splatalogue is a database that includes all known emission and absorption frequencies for atoms and molecules, and it is commonly used for astronomical spectroscopy.
According to Splatalogue, 511 known molecular or atomic species emit or absorb frequencies in X band, and these signals map into 3144 of our spectral channels.
Therefore, these channels were disqualified because we would not be able to differentiate between AMC-NS collision signal and molecular or atomic emission.
Despite being disqualified, we investigated the signals in these channels for completeness.
Of the spectral channels disqualified by the ``Splatalogue Test," 122 of them (associated with 64 species) contained signals greater than 3$\sigma_s$ beyond the mean and 27 (associated with 24 species) were detected 3$\sigma_s$ beyond the median.
No signals beyond $5\sigma_s$ were detected.
This analysis indicates that we may be making moderate signal-to-noise measurements of atoms and molecules, which is expected.


\setlength{\tabcolsep}{5pt}
\renewcommand{\arraystretch}{1.2}
\begin{table}[h]
\centering
\begin{tabular}{c|c|c|c}
Observation Number                & 1      & 2      & 4      \\
\hline
total lines at start              & 26,214 & 39,321 & 39,321 \\
after Band Removal Operation      & 22,938 & 32,768 & 32,768 \\
after Gaussian Width Test         & 22,909 & 32,718 & 32,669 \\
after Excess Kurtosis Test        & 22,878 & 32,333 & 31,905 \\
after Splatalogue Cross Check     & 22,833 & 32,148 & 31,831 \\
\hline
Final Spectra                     & 22,833 & 32,148 & 31,831 \\
\hline
\end{tabular}
\caption{
Cut-flow table showing the number of viable spectral channels remaining after each analysis pipeline test.
Each spectrometer bank nominally has 16,384 channels, but approximately 10\% of the channels were removed from each end, yielding a total of 13,107 active channels per bank.
Observation~1 has fewer channels because we used two spectrometer banks.
Observations 2~\&~3 used three spectrometer banks.
Note that the ``Threshold Test" removes spectral channels from individual integrations.
It does not completely disqualify spectral channels from a given observation, so the ``Threshold Test" is not included in this table.
The ``Final Spectra" row gives the total number of unique spectral channels remaining after disqualification.
Note though that spectral channels from different spectrometer banks overlap.
}
\label{tab:cut_flow_table}
\end{table}


\begin{figure*}[p]
\centering
\includegraphics[width=\textwidth]{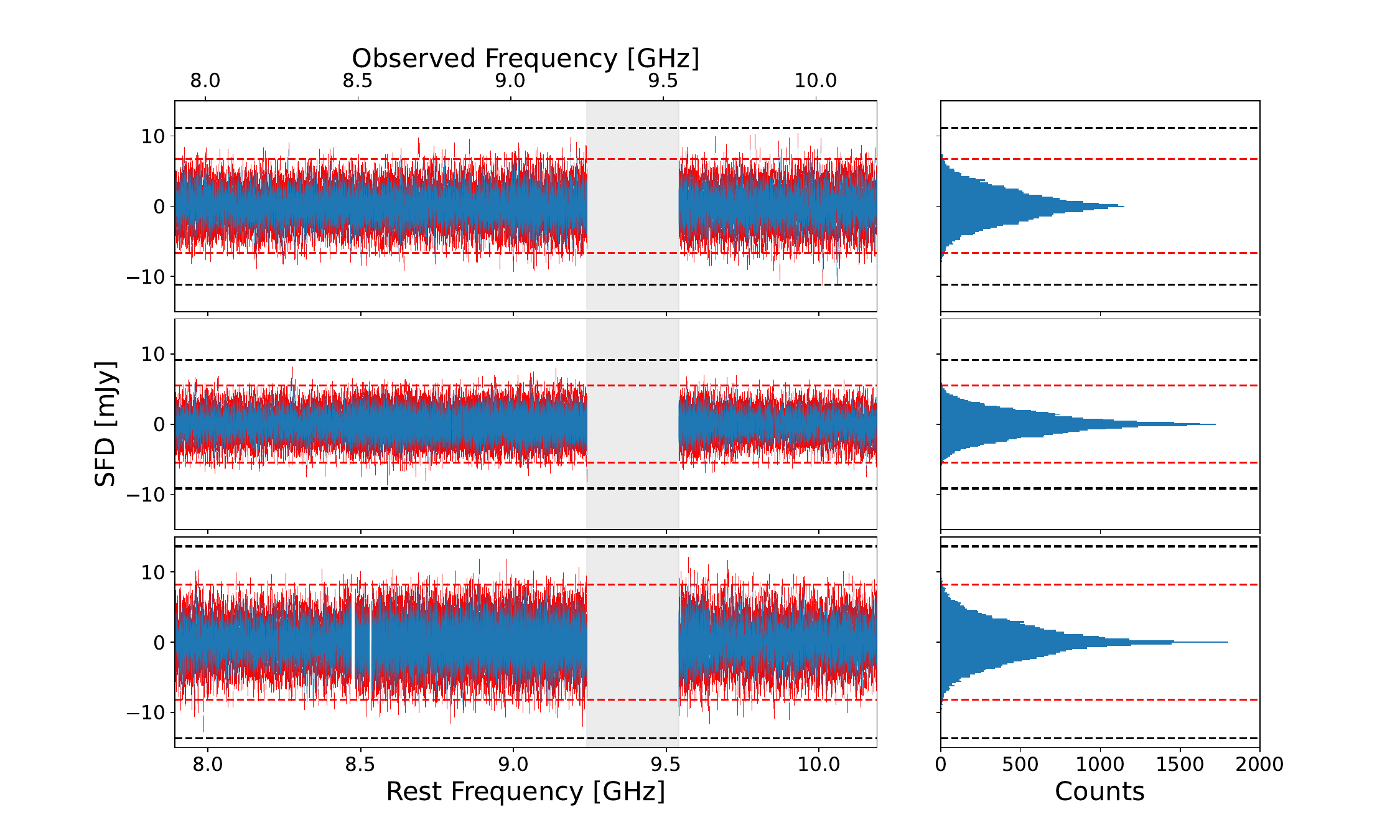} \\
\vspace{-\baselineskip}
\includegraphics[width=\textwidth]{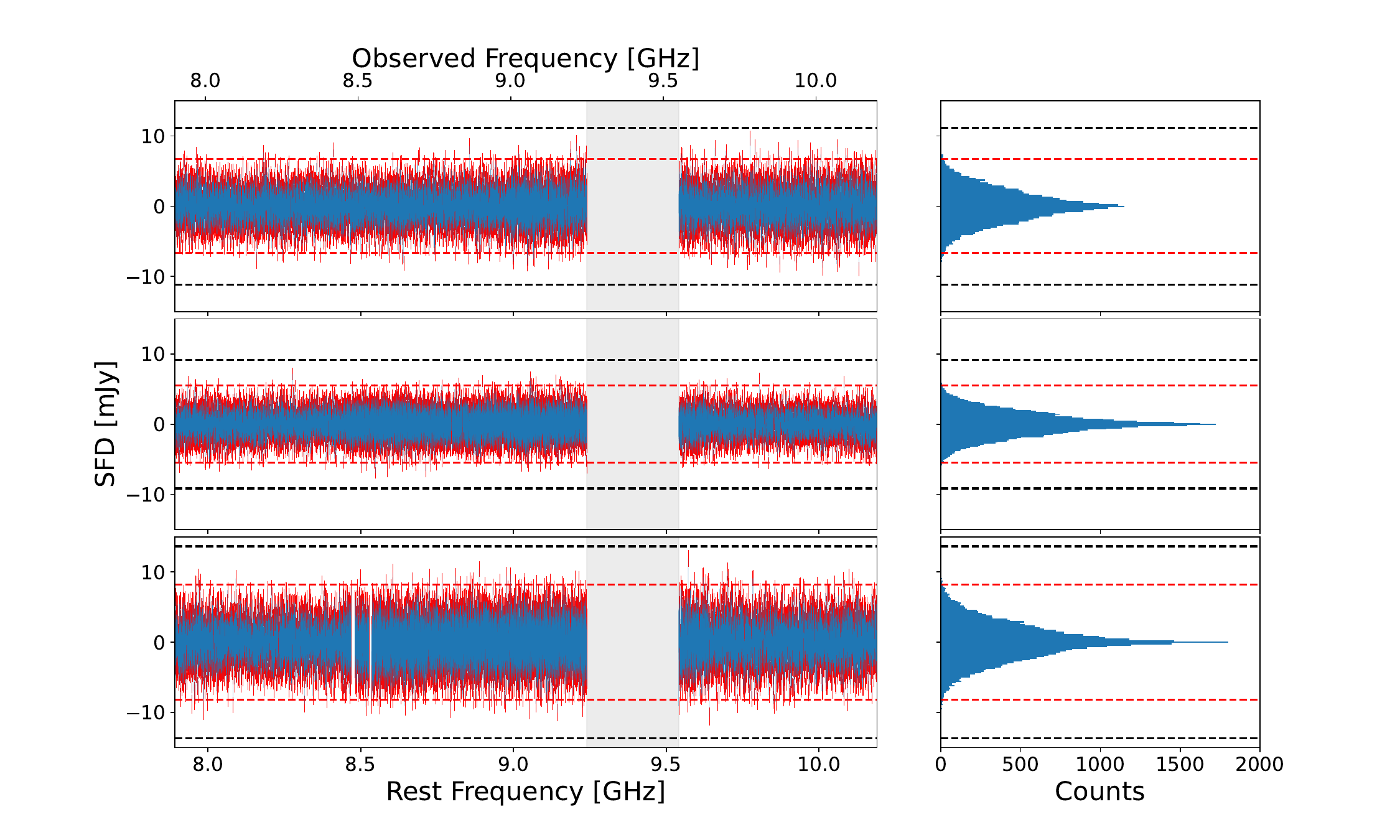}
\caption{
Final median spectra (top) and mean spectra (bottom).
The left sub-panels, from top to bottom, show the final spectra from Observations~1, 2, and 4 after calibration, signal conditioning, and channel disqualification/removal. 
Histograms of the spectra are plotted on the right.
Note that these histograms are somewhat non-Gaussian because of the ``Continuum Smoothing Operation" (see Sec.~\ref{sec:continuum_smoothing}). 
The values for $\sigma_s$ estimated from these histograms are given in Table~\ref{tab:sigma_s}. 
The red and black dashed lines mark $3\sigma_s$ and $5\sigma_s$ detection thresholds. 
See Sec.~\ref{sec:final_spectra} for more detail.
We are looking for an AMC-NS signal in these spectra that is greater than $5\sigma_s$ away from zero in the positive direction.
}
\label{fig:final_spectra}
\end{figure*}


\subsubsection{Final Spectra}
\label{sec:final_spectra}

We computed final mean and median spectra using all scans and integrations in each observation.
Spectral channels disqualified by our analysis pipeline were removed.
A summary of channel disqualification is given in Table~\ref{tab:cut_flow_table}.
The results are shown in Fig.~\ref{fig:final_spectra}. 
Note that some of the channels in the final spectra have data points from multiple spectrometer banks.
The median and mean spectra yield different kinds of information.
Signals that vary on time scales much shorter than the total observation time bias the mean value but not the median value of the spectral channel.
Therefore, signals that appear in the mean spectrum but not in the median spectrum are probably transients with a time scale shorter than 30 minutes.
Signals that appear in both the mean spectrum and the median spectrum are probably stable in time.
We anticipate that the AMC-NS collision signal is largely stable during the two-hour observation, so we are looking for spectral lines that appear in both the mean and median spectra.
In all spectra, the ``Observed Frequency" is the frequency the spectrometer detected, and the ``Rest Frequency" includes a correction we applied, which takes into account the redshift of M31 ($z = -0.001004$), which is moving toward the Milky Way at approximately 300~km\,s$^{-1}$.


\setlength{\tabcolsep}{5pt}
\renewcommand{\arraystretch}{1.2}
\begin{table}[h]
\centering
\begin{tabular}{c|c|c|c}
Observation                          & 1   & 2   & 4 \\
\hline
$\sigma_s$, Mean Spectra [\,mJy\,]   & 2.1 & 1.5 & 2.4 \\
$\sigma_s$, Median Spectra [\,mJy\,] & 2.1 & 1.5 & 2.2 \\
\hline
\end{tabular}
\caption{
The $1\sigma_s$ values for the final spectra shown in Fig.~\ref{fig:final_spectra}.
These results are consistent with sensitivity forecasts (see Sec.~\ref{sec:instrument}).
}
\label{tab:sigma_s}
\end{table}


\setlength{\tabcolsep}{5pt}
\renewcommand{\arraystretch}{1.2}
\begin{table}[h]
\centering
\begin{tabular}{c|c|c|c|c}
        & Observation Number       & 1      & 2      & 4 \\
\hline
        & Spectral Channels        & 22,833 & 32,148 & 31,831 \\
\hline
Mean    & Amplitude $>\,5\sigma_s$ & 0      & 0      & 0 \\
Spectra & Amplitude $>\,3\sigma_s$ & 22     & 6      & 8 \\
\hline
Median  & Amplitude $>\,5\sigma_s$ & 0      & 0      & 0 \\
Spectra & Amplitude $>\,3\sigma_s$ & 27     & 14     & 12 \\
\hline
\end{tabular}
\caption{
Number of candidate spectral channels found in the final spectra.
The ``Spectral Channels" row gives the total number of viable channels for each observation after pipeline disqualification (see Table~\ref{tab:cut_flow_table}).
}
\label{tab:candidate_table}
\end{table}


\subsection{Spectral Channel Analysis}
\label{sec:spectral_channel_analysis}

The AMC-NS collision signal we are searching for should appear in the final mean and median spectra as a single spectral line with an amplitude greater than zero.
Given our observing strategy (see Sec.~\ref{sec:observation_details}) and our pipeline design (see Sec.~\ref{sec:baseline_alignment}~\&~\ref{sec:continuum_smoothing}), positive spectral features are produced by emission signals originating in M31, while negative spectral features originate from our off-target pointing location, or they are produced by molecular or atomic absorption.
By eye, the final mean and median spectra are largely noise.
If the spectra contain only Gaussian noise, then we expect to find
\begin{align}
N = 0.8 \, \alpha \, N_\mathrm{ch} \, N_\mathrm{b} \, N_\mathrm{obs}
\end{align}
channels with signal greater than $3 \sigma_s$ or $5 \sigma_s$.
Here, $N_{ch}$ is the number of spectral channels per spectral bank, $N_{b}$ is the number of spectral banks, $N_{obs}$ is the number of observations, and $\alpha = 3 \times 10^{-3}$ for $3\sigma_s$ and $\alpha = 6 \times 10^{-7}$ for $5\sigma_s$.
We estimated $\sigma_s$ by fitting a one-dimensional Gaussian to the histograms shown in Fig.~\ref{fig:final_spectra} using the method discussed in Sec.~\ref{sec:spectral_channel_noise_test}.
The estimated values for $\sigma_s$ are given in Table~\ref{tab:sigma_s}.
The value of $0.8$ is included because we cut 20\% of the channels in each spectral bank to account for filter drop-off (10\% from each end). 
Assuming we have 16,384 channels per spectral bank, three banks for two observations, and two banks for one observation, we expect to have at most 315 spectral channels with a signal amplitude beyond 3$\sigma_s$ and zero beyond 5$\sigma_s$.
%
%
Our actual result is summarized in Table~\ref{tab:candidate_table}.



\begin{figure}
\includegraphics[scale=0.2]{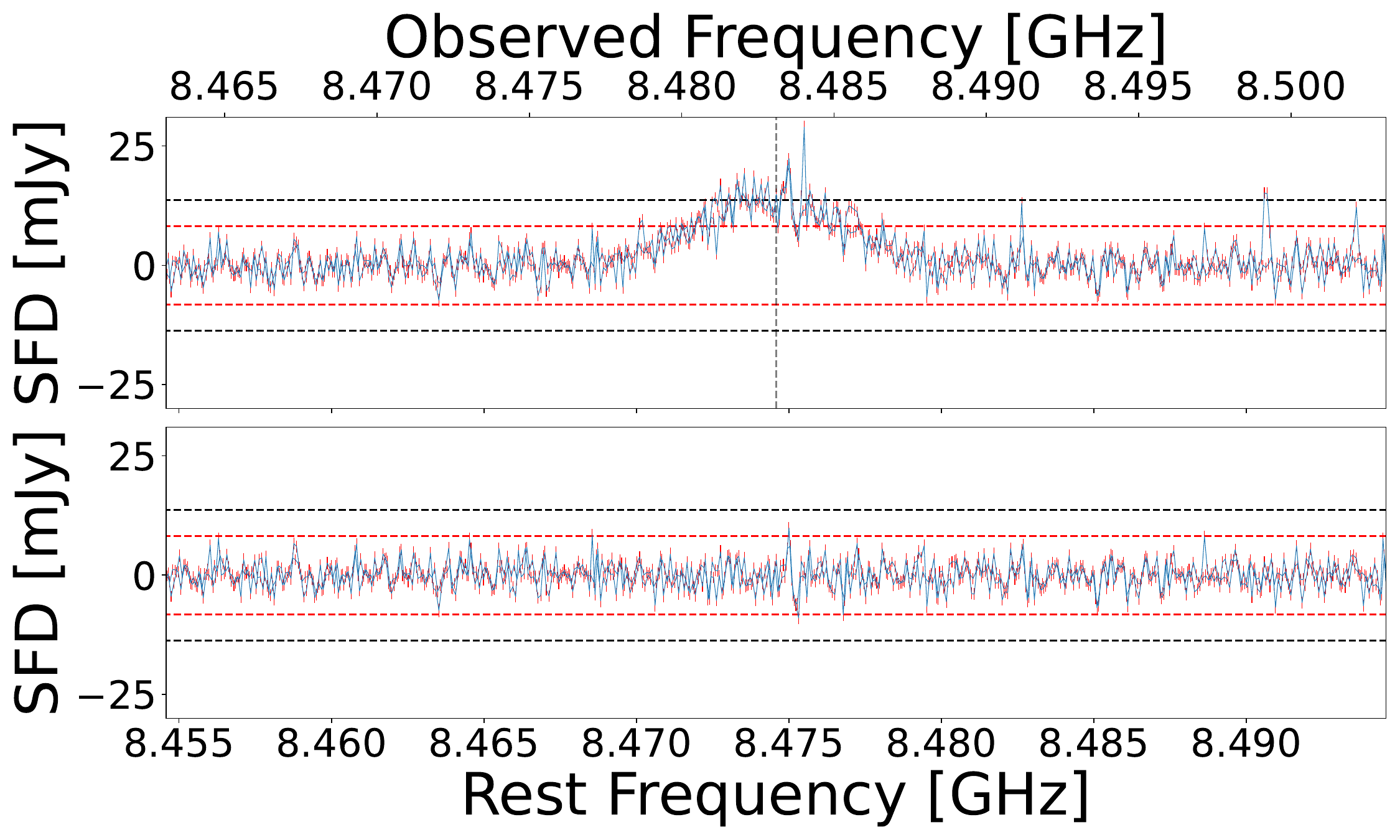} \\
\vspace{0\baselineskip}
\includegraphics[scale=0.2]{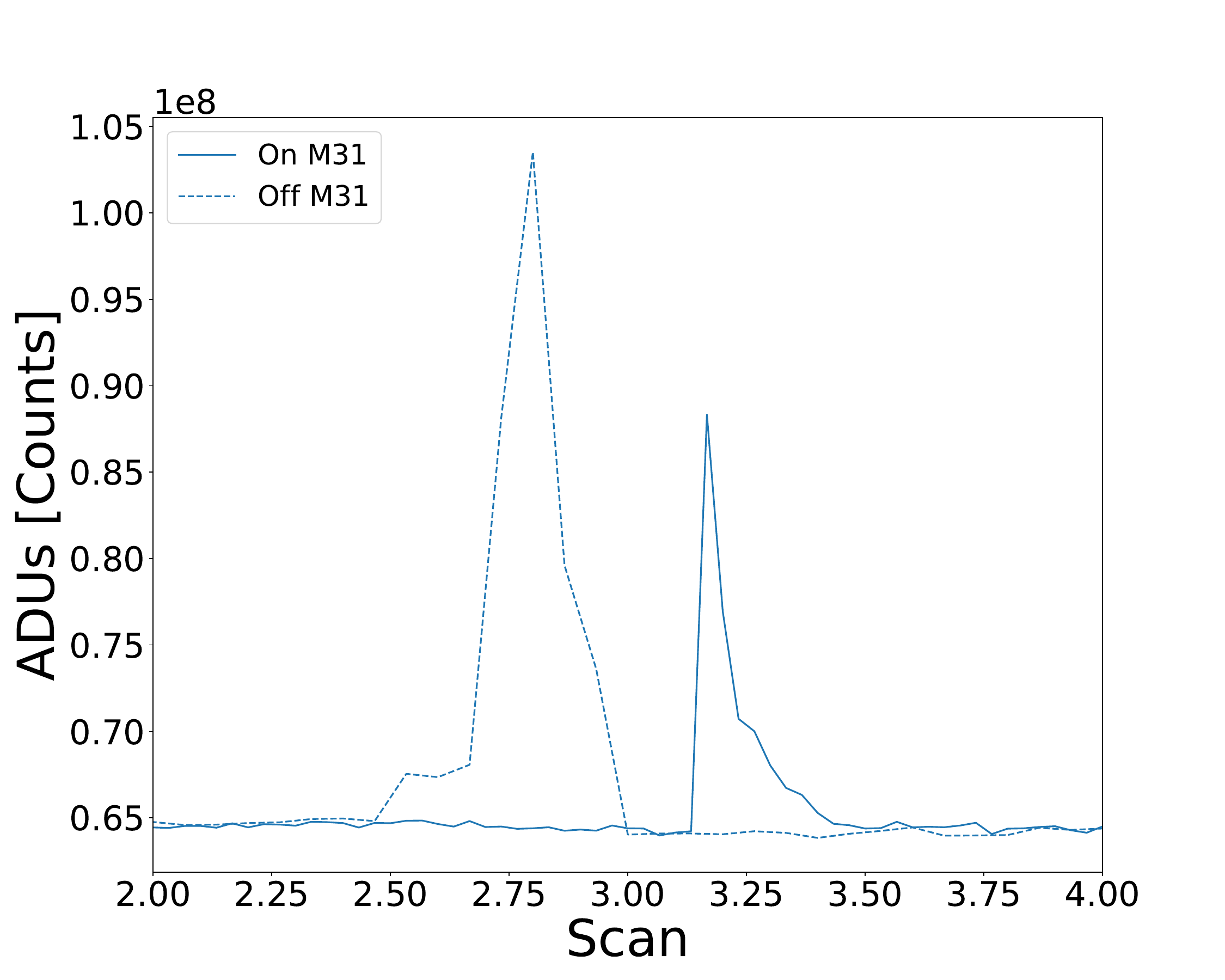}
\caption{
An example signal from Observation~4 that was rejected by our pipeline.
In the top panel, the blue line shows the mean (top sub-panel) and median (bottom sub-panel) spectra, and the red lines show the uncertainty ($\sigma_{\nu}$).
The horizontal dashed lines mark our $3\sigma_s$ (red) and $5\sigma_s$ (black) detection limits. 
The bottom panel shows the spectral channel data plotted versus scan number, which is a proxy for time.
The frequency of the particular feature shown matches a hydrogen recombination line H($\rm131\gamma$) at 8.483 GHz when we do not account for the blueshift of M31. 
Therefore, if it is hydrogen recombination line emission, then it would be from a cloud in the Milky Way.
When we do account for blueshift, the closest spectral line to the center of the feature is bromine dioxide (8.4922~GHz), which is not a common astrophysical molecule~\cite{mcguire_2022}.
Note that the spectral breadth of this signal suggests it does not originate from an AMC-NS collision. 
See Sec.~\ref{sec:spectral_feature_examples} for more detail.
}
\label{fig:hgamma}
\end{figure}


\begin{figure}
\includegraphics[scale=0.2]{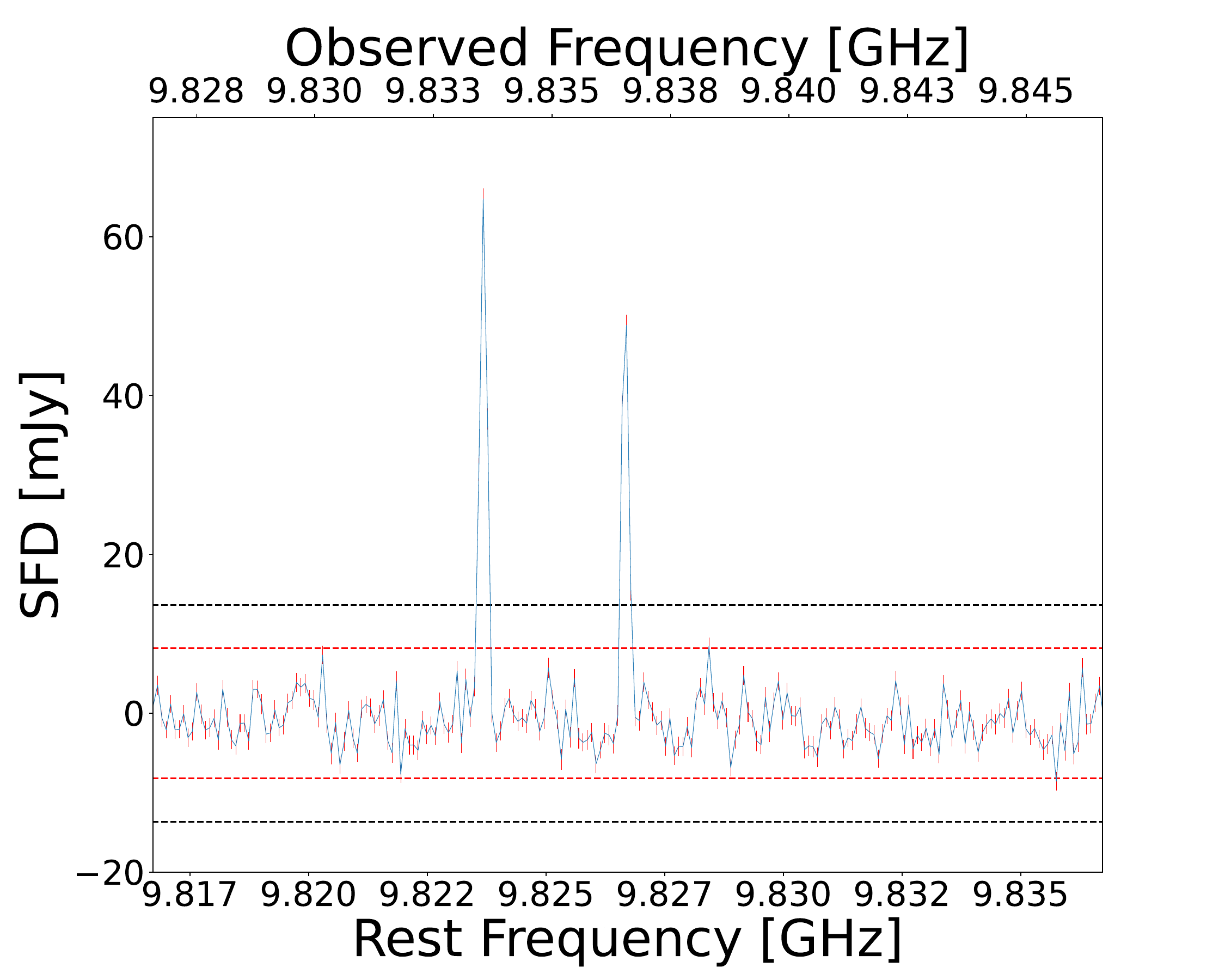} \\
\vspace{0\baselineskip}
\includegraphics[scale=0.2]{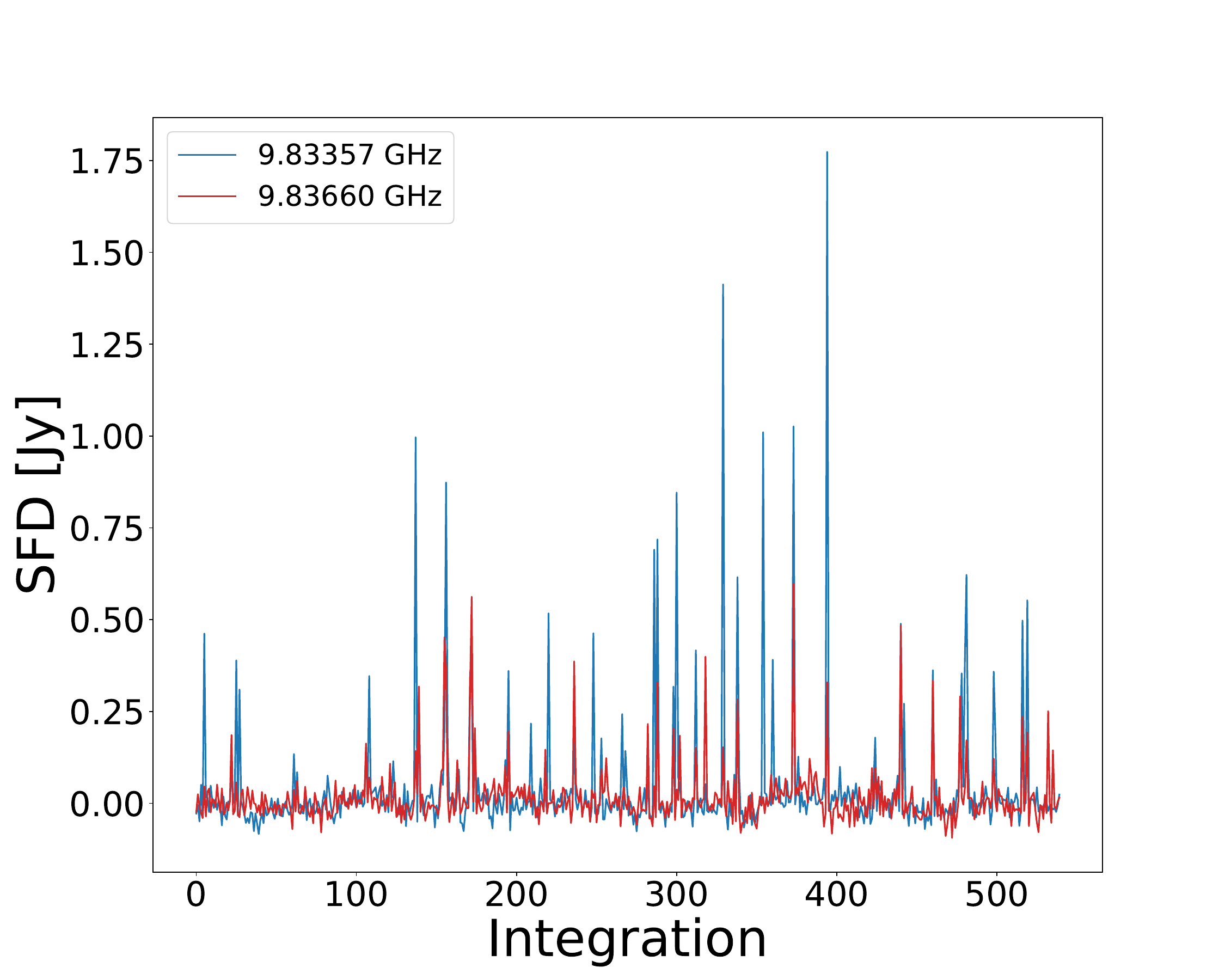}
\caption{
Another example signal from Observation~4 that was rejected by our pipeline.
In the top panel, the blue line shows the mean spectrum, and the red line shows the uncertainty.
The horizontal dashed lines mark our $3\sigma$ (red) and $5\sigma$ (black) detection limits. 
The frequencies associated with the two narrow lines match molecular transitions of ammonia (9.833~GHz) and $\rm \alpha$-alanine (9.8366~GHz) when we do not account for blueshift. 
These signals are best seen in our fourth observation (shown) but can also be seen as fainter signals in our other observations. 
When we do account for blueshift, we do not find any known atomic or molecular lines that match, which means they could be candidate axion emission lines. 
However, we also detect other known $\rm \alpha$-alanine lines less than $5\sigma_s$, which means $\rm \alpha$-alanine emission in the Milky Way is a possible explanation. 
The bottom panel shows the amplitude of these lines versus integration number, which is a proxy for time.
The computed excess kurtosis values for these channels are 38 and 18, and the $1\sigma_{\nu}$ noise levels are 31 and 32~mJy.
Note that these signals are intermittent, which suggests RFI is probably the best explanation.
See Sec.~\ref{sec:spectral_feature_examples} for more detail.
}
\label{fig:ammonia}
\end{figure}


\subsection{Spectral Feature Examples}
\label{sec:spectral_feature_examples}


Figs.~\ref{fig:hgamma}~\&~\ref{fig:ammonia} show a few example spectral features that our pipeline discovered and rejected as candidate AMC-NS collision signals.
The example signals shown were detected during Observation~4. 
The broad feature that matches a hydrogen recombination line (H131$\rm \gamma$) only appears in Observation~4. 
This implies that some time variable phenomenon occurred during this observation on a time scale shorter than the observation.
One possibility is a hydrogen cloud in the Milky Way passing through our beam. 
The lines that look like ammonia and alanine are both apparent in the data from Observation~1~\&~2 as well, but the signal just barely breaks the 5$\sigma_s$ threshold.
While these are not the AMC-NS collision signals we are searching for, they demonstrate that our pipeline is effective and capable of finding similar signals in the data.


\section{Discussion}
\label{sec:discussion}


\subsection{Interpretation of our Results}
\label{sec:interpretation}

We chose to observe in X band because in August 2021, when we proposed our first observations, the models of \citet{Edwards:2020afl} and \citet{Witte:2021arp} suggested that in M31, each day there should be $\mathcal{O}(1)$ AMC-NS collisions that would produce a signal above 1~mJy around 10~GHz.
The theoretical modeling has matured since August 2021.
We now find a peak event rate near 3~GHz, and we find that detectable events in X band should be rare, requiring longer and wider observations to catch an AMC-NS collision\footnote{The predicted peak in event rate at 3~GHz in our current model raises an interesting dichotomy for AMC-NS searches: the event rate only becomes high at particle masses below the expected range for the QCD axion in the cosmology that gives rise to AMCs.
AMCs made from ALPs should be the target at lower frequencies, where the neutron star population model predicts a more substantial signal (see \citet{OHare:2021zrq} and \citet{Gorghetto:2022ikz})}.
The null result from the short observations reported in Sec.~\ref{sec:results} are thus consistent with our best current model for AMC and neutron star populations. 
Therefore, the null result does not set a limit in the axion parameter space $(m_a,g_{a\gamma\gamma})$.
In any case, setting a limit would depend on the multi-parameter population model for AMCs and neutron stars.
With such a short total observing time, and observing only M31, a multi-parameter constraint accounting for population uncertainties would not be enlightening (more in Sec.~\ref{sec:future_experiment}).


\subsection{Additional Observations}
\label{sec:additional_observations}

We are continuing the search for AMC-NS collisions in M31 in other spectral bands, targeting the aforementioned event-rate peak near 3~GHz.
In 2023, we completed follow-up observations of M31 using the C-band receiver with the VEGAS spectrometer at the GBT (award number GBT23A-245).
These observations probe 4.0 to 8.0~GHz, which corresponds to the axion mass range 16 to 33~$\mu$eV.
We were also awarded observing time with the ultra-wideband (UWB) receiver, which can observe between 0.7 and 4~GHz.
However, we did not collect any data because of a technical problem with the instrument that semester.
We plan on making these UWB observations in the future.
Currently, we are making observations with the L-band receiver/spectrometer on the 20~m Telescope at the GBO.
These observations probe 1.3 to 1.8~GHz, which corresponds to the axion mass range 5.4 to 7.4~$\mu$eV.
The analysis of all additional data is ongoing.
The analysis pipeline presented in this paper for our X-band observations will be reused on these additional observations.

As a separate line of inquiry, in 2023 we were awarded observing time on the ARO 12-meter Telescope to make preliminary observations of a single neutron star RBS1223 (one of the ``Magnificent Seven"~\cite{haberl_2007}).
For these observations, we used the 2 and 3~mm receivers with the wideband spectrometer (AROWS) at maximum bandwidth (4~GHz), which gives $\sim$600~kHz spectral resolution.
We chose to observe RBS1223  because it is young (1.5~Myr) and nearby, which should help with signal brightness.
These observations probe 84 to 95~GHz and 140 to 158~GHz, which corresponds to a much higher axion mass range 350 to 650~$\mu$eV~(see~\citet{Aja:2022csb} for a discussion of a laboratory-based experiment probing this mass range).


\begin{figure*}
\centering
\includegraphics[width=0.75\textwidth]{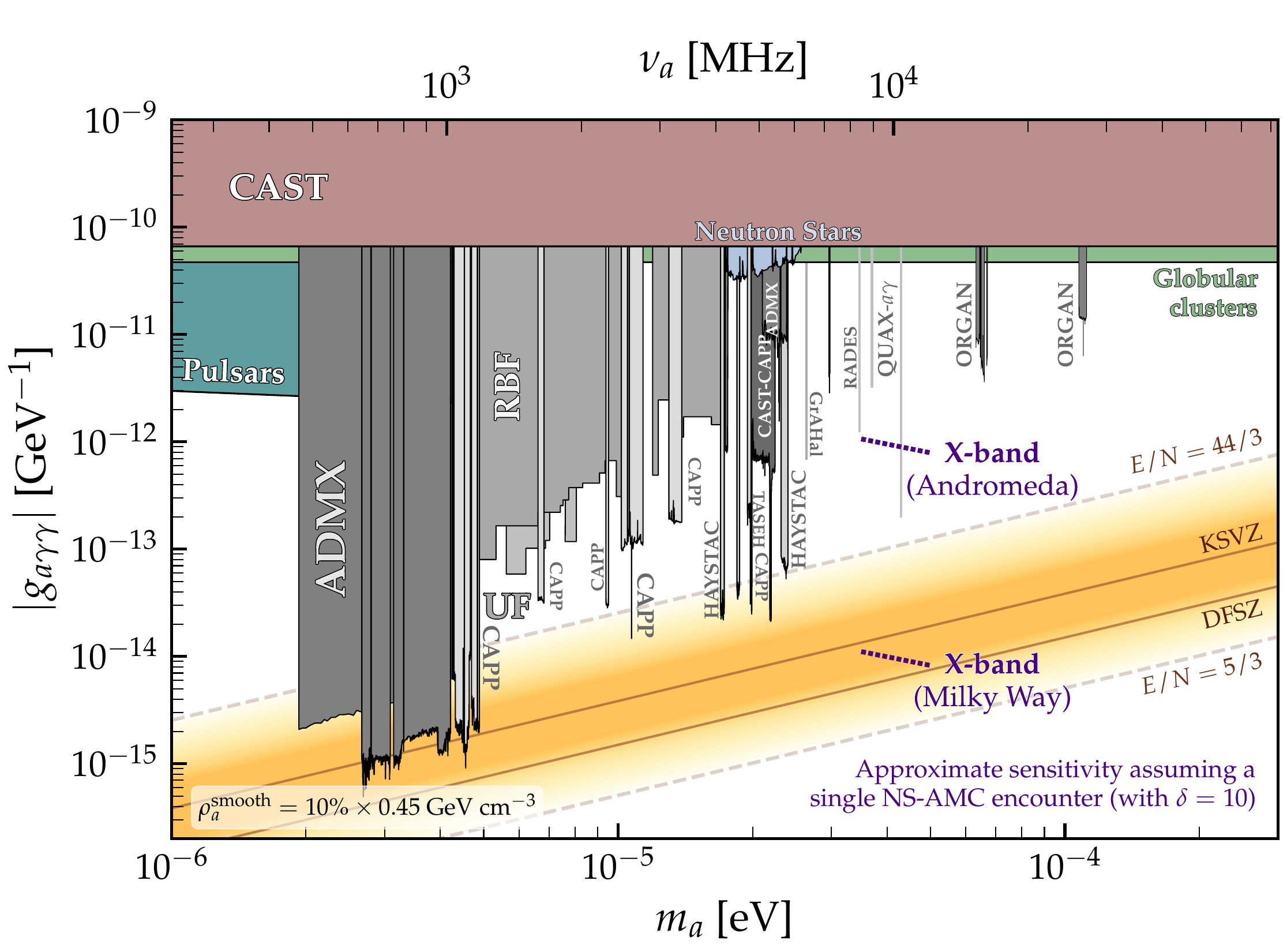}
\caption{
Estimated sensitivity to axion-photon couplings assuming the observation of a single AMC-NS encounter. 
Sensitivity is estimated by calculating the value of $g_{a\gamma\gamma}$ required to achieve a mean flux density during the encounter of 2~mJy for events in Adnromeda (where only AMC transients would be bright enough) and the Milky Way (which requires a larger search area, but has continuous and transient signals).
We assume an AMC with a mass $M = 10^{-10}\,M_\odot$ and an overdensity of $\delta = 10$. 
We assume that the neutron star has a polar B-field $B_0 = 10^{14}\,\mathrm{G}$ and a rotation period $P = 1\,\mathrm{s}$. 
The encounter takes place with an impact parameter $b = 2 R_\mathrm{AMC}/3$ (the average from our model). 
The rate of encounters depends on detailed modeling of the AMC and neutron star populations.
The precise scan strategy and amount of observing time needed to achieve this result is something we are investigating now.
For comparison, we show the QCD axion model band (yellow) and a number of existing constraints from haloscopes, helioscopes, pulsars (vacuum gap production), neutron stars (continuous signal), and globular clusters (see~\citet{AxionLimits} and references therein).
For the limits from haloscopes (grey) and neutron stars (pale blue), we assume that the smooth axion distribution makes up only 10\% of the total dark matter density, consistent with the AMC scenario~\cite{Eggemeier:2022hqa}.
This figure is adapted from \href{https://cajohare.github.io/AxionLimits/}{AxionLimits}~\cite{AxionLimits}.}
\label{fig:AxionLimits}
\end{figure*}


\subsection{Possible Future Experiment}
\label{sec:future_experiment}

Our main takeaway from this study is that searching for AMC-NS collisions with radio telescopes is promising.
The needed high-resolution spectrometer technology is already available and it can be deployed across a broad range of frequencies in the microwave and millimeter-wave bands, so measuring $m_a$ is straightforward for radio telescopes if the AMC-NS collision signal actually exists and it is bright enough to be detected.
A measurement of $g_{a\gamma\gamma}$ with radio telescopes, however, will be challenging for now because accurate signal modeling is needed, which in turn requires a better understanding of the neutron stars.
The radio telescope approach nicely complements laboratory-based searches.
The axions in AMCs should be the same as the axions in the diffuse dark matter near Earth.
Therefore, the discovery of an AMC-NS collision signal -- or even a candidate signal -- would usefully inform laboratory-based experiments.
One possible way to proceed is to exploit the more nimble nature of radio telescopes and use them to search broadly for $m_a$.
Then tailor haloscopes for follow-up measurements of both $m_a$ and $g_{a\gamma\gamma}$ if a candidate mass is detected.

We also learned that state-of-the-art radio telescopes are not necessarily optimized for blind axion searches for two primary reasons.
First, a significant amount of observing time is needed to integrate down the noise.
Observing with the GBT for months or years is impossible because it is a shared resource. 
Observing time awards are commonly hours or days.
Second, a small telescope beam is ideal for detecting faint radio sources and running morphological studies of astrophysical objects, but it is not optimal for trawling the galaxy for young, high-$B$ neutron stars.
Neutron stars are very small with respect to other astronomical objects, having a radius $R_{\rm NS} \sim$ 10\,km.
Therefore, despite the fact that they are are hot, their luminosity is small when compared with main sequence stars.
Consequently, neutron stars are hard to detect at visible wavelengths, so we do not know precisely where they all are.
Nevertheless, we do know there are a large number of them in the Universe because they are dead, high-mass, main-sequence stars.
To catch as many AMC-NS collisions as possible, blind searches in regions where suitable neutron stars might be is appropriate.

This all suggests that a dedicated radio telescope could be helpful.
This instrument should (i) target neutron star populations in the Milky Way, (ii) survey (search and monitor) a large fraction of the sky, so many nearby neutron stars can be simultaneously observed, increasing the probability of detecting a signal, (iii) have a large beam (small-diameter reflector), and (iv) synchronously observe with an omnidirectional antenna capable of detecting and monitoring RFI.
AMC-NS collision signals originating in the Milky Way would be brighter because the sources would be orders of magnitude closer.
Observing bright signals would allow access to lower values of $g_{a\gamma\gamma}$ and $\delta$.
We chose to observe M31 because a large number of (distant) neutron stars could be observed with a small amount of observing time.
To observe a similarly large number of nearby neutron stars in the Milky Way, a larger portion of the sky would have to observed.
A dedicated radio telescope would enable the needed long observing sessions (months or years) that are not realistically available at the GBT. 
The recent high resolution cosmic string simulations~\cite{Buschmann:2021sdq} suggest 10 to 40~GHz is a theoretically motivated microwave band to search.
The upper portion of this mass range will be more challenging because of the rarity of high-$B$ neutron stars and atmospheric opacity becomes a possible issue.
Therefore, X and Ku bands (the lower end of the mass range) are likely the best bands to start with.

In Fig.~\ref{fig:AxionLimits}, we illustrate the possible reach in $(m_a,g_{a\gamma\gamma})$ space of a dedicated radio telescope operating in X-band.
This plot assumes the instrument observes for long enough and over enough of the sky that AMC-NS population models predict a single event for a given $(B_0,\delta)$ combination, either in the Milky Way or Andromeda.
The precise scan strategy and amount of observing time needed to reach these limits is something we are investigating now, but we think the possibility of probing the QCD axion in parts of the natural parameter space in the post-inflation scenario motivates significant future effort in this direction.


\subsection{Additional Thoughts on AMC Studies}
\label{sec:additional_thoughts}

Interpreting results, or null results in this context, still requires significant work in understanding the AMC and neutron star populations.
Until a better understanding is reached, what would the implications of a detection be?
How could we identify a candidate radio emission line as being due to an AMC-NS collision? 
The axion line is very narrow, so we could get no spectral information about it.
If a line was found, returning to it to gain more time series information could be useful to pin down aspects of the AMC distribution, since AMCs of different masses and density profiles would typically have collisions of different duration~\cite{Edwards:2020afl}.
Due to the dependence of the signal on the neutron star magnetosphere itself~\cite{Witte:2021arp,McDonald:2023shx} (expressed in simple terms in Equation~\ref{eq:power}), from a signal one could also infer some properties of the neutron star, for example, that it must have a particular magnetic field or inclination.

From the point of view of axion detection the most immediate theoretical implication is that the axion mass would be uniquely determined by the line frequency.
Follow up by direct detection haloscopes~\cite{Sikivie:1983ip}, particularly in X band and below, is possible with existing technology.
For example, had we detected a line in X band, the haloscopes CAST-RADES~\cite{Melcon:2018dba,CAST:2020rlf}, QUAX~\cite{Alesini:2022lnp}, and the upcoming ALPHA~\cite{ALPHA:2022rxj} would be able to tune to this frequency and find the same line caused by the halo of the MW.
In the AMC model, a halo based search is not generally sensitive to the AMCs themselves due to their rarity (although see \citet{Tinyakov:2015cgg,OHare:2017yze,OHare:2023rtm}), however a minimal target is provided by the complement: `minivoids'~\cite{Eggemeier:2022hqa}.
For this reason, in the AMC scenario, frequencies already probed and nominally excluded by haloscopes still need to be searched for transient signals, since haloscopes currently do not scan deep enough to exclude the QCD axion in the AMC scenario.
A transient signal would motivate such a deep rescan.
Moreover, while the technology to search for galactic axions in this band is already deployed in laboratories, the time required to scan a given mass range can last several months, especially if the weaker signal at the minivoid density must be reached. The transient search also takes significant time to reach an expected number of events of order unity, but has the advantage of being broadband.

If a transient line was detected, others might appear in different places where AMC-NS collisions might occur frequently.
Subsequent lines at the same frequency with different time-domain properties in different locations could then be used to cement the relationship between the AMC and neutron star populations and the signal properties.
In the event of detection of a candidate AMC-NS line, a combination of astrophysics and direct detection could thus pin down not just the axion mass and coupling, but also minute details of the dark matter substructure, and the neutron star population.


\section{Conclusions}
\label{sec:conclusions}


We conducted four two-hour observations of the center of the Andromeda galaxy in X band using the GBT to perform the first dedicated search for the radio transient signal caused by the collision of an AMC with a neutron star magnetosphere.
After calibrating our spectra and taking appropriate data cuts, we found seven candidate lines surpassing the $5\sigma_s$ detection threshold.
Inspecting each of these signals in the time and frequency domains, none match the expected signal.
We conclude that no detectable AMC-NS transient occurred in the observed region of M31 during our observations. 
With our current understanding of AMC and neutron star populations it is difficult to use this null result, with a short observation time, to set a limit on axion parameters in X band.
However, with the experience we have gained we propose a dedicated observing facility to continue this search.
A radio line from M31 could only be caused by the relatively loud AMC-NS collision signal, and with sufficient observing time to overcome the rarity of events, M31 observations can place relevant limits on axion parameter space just above the predicted QCD model band.
A radio transient in the Milky Way, on the other hand, is also sensitive within the QCD model band.

In this study we learned some hard truths about fishing for AMC-NS transients.
Nonetheless, we are confident that such transients offer an opportunity to search for the QCD axion and ALPs in well-motivated parameter space.
The difficulties of realising this promise are significant, requiring possibly years of dedicated observation and complex modeling of dark matter substructure and galactic neutron star populations.
We hope our first foray into this heretofore uncharted territory inspires this necessary future work.


\begin{acknowledgments}

We would like to thank the Green Bank Observatory for providing the observing time on the GBT (award GBT22A-067).
We would like to thank the GBO staff; in particular, Dave Frayer, Brenne Gregory, and Jesse Bublitz who helped us make the observations, and Steve White and Jason Ray, who explained the receiver and the VEGAS hardware.
Tom Edwards, Scott Ransom, and Sam Witte provided valuable expertise when setting up this project, and we would like to acknowledge their helpful and important contributions.
We would also like to thank Christoph Weniger for helping us get the project started. 
L.W.\ thanks Amanda Alvarado for her help with identifying potential signals in our data.
M.E.\ is supported by an NSF/GRFP fellowship.
B.J.K.\ acknowledges funding from the Ram\'on y Cajal Grant RYC2021-034757-I, financed by MCIN/AEI/10.13039/501100011033 and by the European Union ``NextGenerationEU''/PRTR.
B.J.K.\ also acknowledges support from the DMpheno2lab Project (Project PID2022-139494NB-I00 financed by MCIN /AEI /10.13039/501100011033 / FEDER, UE).
We acknowledge the Santander Supercomputation support group at the University of Cantabria who provided access to the Altamira Supercomputer at the Institute of Physics of Cantabria (IFCA-CSIC), member of the Spanish Supercomputing Network, for performing simulations/analyses.
D.J.E.M.\ is supported by an Ernest Rutherford Fellowship from the STFC, Grant No. ST/T004037/1.
L.V.\ acknowledges support by the National Natural Science Foundation of China (NSFC) through the grant No.\ 12350610240 ``Astrophysical Axion Laboratories''.

\end{acknowledgments}


\raggedright
\bibliography{apssamp}
\end{document}